\documentclass[12pt]{iopart}
\usepackage{amssymb}
\usepackage{dcolumn} 
\usepackage{graphicx}
\usepackage{hyperref}
\usepackage{color}
\usepackage{colortbl}
\usepackage[tabel]{xcolor}
\usepackage{braket}
\usepackage{multirow}
\usepackage{subfigure}
\usepackage{bm}

\renewcommand{\t}[1]{\textrm{\scriptsize #1}}
\renewcommand{\vec}[1]{\mathbf{#1}}

\begin{document}
\title[All-Electron, Real-Space Perturbation Theory for Homogeneous Electric Fields]{All-Electron, Real-Space Perturbation Theory for Homogeneous Electric Fields: Theory, Implementation, and Application within DFT}

\author{Honghui Shang$^1$, Nathaniel Raimbault$^1$, Patrick Rinke$^2$, Matthias Scheffler$^1$, Mariana Rossi$^1$, and Christian Carbogno$^1$}

\address{$^1$ Fritz-Haber-Institut der Max-Planck-Gesellschaft, Faradayweg 4--6, D-14195 Berlin, Germany}
\address{$^2$ COMP/Department of Applied Physics, Aalto University, P.O. Box 11100, Aalto FI-00076, Finland}

\eads{\mailto{carbogno@fhi-berlin.mpg.de}, \mailto{rossi@fhi-berlin.mpg.de}}

\begin{abstract}
Within density-functional theory, perturbation theory~(PT) is the state-of-the-art formalism for assessing the response to homogeneous electric fields and the associated material properties, e.g., polarizabilities, dielectric constants, and Raman intensities. Here, we derive a real-space formulation of PT and present an implementation within the all-electron, numeric atom-centered orbitals electronic structure code FHI-aims that allows for massively-parallel calculations. As demonstrated by extensive validation, we achieve a rapid computation of accurate response properties of molecules and solids. As an application showcase, we present harmonic and anharmonic Raman spectra, the latter obtained by combining hundreds of thousands of PT 
calculations with \textit{ab initio} molecular dynamics. By using the PBE exchange-correlation functional 
with many-body van der Waals corrections, we obtain spectra in good agreement with experiment especially with respect to lineshapes for the isolated paracetamol molecule and two polymorphs of the paracetamol crystal.
\end{abstract}

\vspace{2pc}

\submitto{\NJP}

\maketitle

\section{Introduction}
The response of molecules and solids to an applied electric field is a fundamental physical mechanism of prime importance, since it determines significant properties and 
spectroscopic signals, such as dielectric constants, Raman spectra, and sum-frequency generation spectra. In first-principles frameworks, these quantities are typically
computed via time-dependent density-functional theory~\cite{DalCorso1996,Andrade:2007kg} or  via analytical perturbation theory~(PT) in either its density-functional perturbation theory~(DFPT)\cite{Gonze1997-1,Gonze1997-2, PutrinoParrinelloDFPT2000, Baroni-2001} 
or coupled perturbed self-consistent field~(CPSCF) formulation~\cite{Gerratt-1967,Pople-1979,Dykstra-1984,Frisch-1990, Ochsenfeld-1997, Liang-2005}. 
Within these linear-response approaches, an additional complexity arises for the treatment of solids: As discussed in more detail in Sec.~\ref{ssec:solids}, the position operator appearing in the respective
equations is not well defined and the expressions need to be recast into a more suitable form.  Practical implementations of these methods within Kohn-Sham density-functional 
theory (DFT) differ substantially by their choice of basis sets (e.g.,~plane-waves or localized basis sets) and by their treatment of the core electrons~(e.g.,~all-electron 
or pseudopotentials). In this paper, we address an implementation of PT for the response to a homogeneous electric field targeted towards handling large periodic systems, which will be the subject of our showcase in Sec. \ref{sec:application}. Our implementation uses the all-electron, numeric atom-centered orbital based framework of the FHI-aims code \cite{Blum2009,Havu/etal:2009,Ren/etal:2012}, which also features implementations of PT for vibrational~\cite{Shang:2017hn} and magnetic properties~\cite{Laasner:fx}. Notably, this infrastructure allows us to treat isolated systems (such
as molecules) and extended systems (such as periodic, crystalline solids) on the same footing, as discussed in Sec.~\ref{sec:theory} and \ref{ssec:solids}, respectively.

As an application, we focus on the important task of calculating anharmonic vibrational Raman spectra of molecular crystals. These spectra are able to provide information about differences in the polymorphic structure of these crystals, the presence of impurities, and the onset of phase transitions. Importantly, they are quantities that can be readily accessed experimentally under different thermodynamic conditions, which can also be unambiguously simulated. In that respect, the calculation of these spectra in an anharmonic fashion using time-correlation functions~\cite{Putrino_PRL_2002,Pagliai_JCP2008}, as further detailed in Sec. \ref{sec:application}, represents an important link between computer simulations and experiments. It can help to gauge the impact of anharmonicities in different phonon modes, which opens the path for a better understanding and control of the polymorphic forms of molecular crystals. The particular character of our implementation results in an efficient scaling with respect to system size (due to a sparse representation of the density matrices) and efficient numerical scaling with respect to the number of cores used on modern, massively parallel architectures~(due to the use of local real-space operations). This facilitates the calculations of tens of thousands of polarizability tensors along \emph{ab initio} molecular dynamics~(MD) trajectories and thus enables the evaluation of anharmonic Raman spectra. We discuss how these spectra depend on different functionals and propose ways to obtain them at minimal cost. Our \textit{ab initio} spectra computed at room temperature compare very well with experimental data obtained at the same conditions.

The remainder of this paper is organized as follows. The fundamental perturbation theory framework is discussed for finite and extended periodic systems in Sec.~\ref{sec:theory} and \ref{ssec:solids}, respectively. In Sec.~\ref{sec:Implementation}, a detailed derivation of the respective equations and their implementation in a real-space, all-electron, numeric atom-centered orbitals based framework is presented. In Sec.~\ref{sec:results}, our approach and implementation is validated by comparing the calculated analytical polarizabilities and dielectric constants to literature values or to ones computed via finite-differences. Furthermore, we discuss the convergence behavior of our implementation, the scaling with system size, and the parallel performance when a large number of cores is used. We finish in Sec.~\ref{sec:application} by applying the developed formalism to compute harmonic and anharmonic Raman spectra for
different polymorphs of the paracetamol crystal.

\section{Fundamental Theoretical Framework}
\label{sec:theory}
Before addressing the implementation in the FHI-aims code, we recall the basic equations used in this work. Throughout the text, we use a spin-unpolarized notation for the sake of simplicity, but a formal generalization to a collinear (scalar) spin treatment is straightforward. Moreover, we focus on systems with a non-vanishing energy gap for electronic excitations, because electric fields in metals are fully screened~(Our  numerical strategy to deal with quasi-degenerate electronic states in non-metallic systems is discussed in Sec.~\ref{sec:C1}.). In this section, a detailed derivation of the equations for finite, molecular systems is given; a generalization to extended periodic solids follows in Sec.~\ref{ssec:solids}.

In Kohn-Sham DFT, the total-energy functional is given by 
\begin{equation}
E_\t{KS}[n]= T_\t{s}[n]+E_\t{ext}[n]+E_\t{H}[n]+E_\t{xc}[n] + E_\t{nuc-nuc}\;.
\label{eq:KSTOT}
\end{equation}
Here, $n(\mathbf{r})$ is the electron density, $T_\t{s}$ the kinetic energy of non-interacting electrons,  $E_\t{ext}$ the external energy due to the electron-nuclear attraction, $E_\t{H}$ the Hartree energy, $E_\t{xc}$ the exchange-correlation energy, and $E_\t{nuc-nuc}$ the repulsion energy of the nuclei.
The ground state electron density~$n_0(\vec{r})$ (and the associated ground-state total energy) 
is obtained by variationally minimizing Eq.~(\ref{eq:KSTOT}) under the constraint that
the number of electrons $N_e$ is conserved. This yields the chemical potential $\mu=\delta E_\t{KS}/\delta n$ of the electrons
and the Kohn-Sham single particle equations
\begin{equation}
\hat{h}_\t{KS}\psi_p = \left[ \hat{t}_\t{s} + \hat{v}_\t{ext}+\hat{v}_\t{H}+\hat{v}_\t{xc}\right] \psi_p = \epsilon_{p} \psi_p 
\label{eq:ks-equation}
\end{equation}
for the Kohn-Sham Hamiltonian~$\hat{h}_\t{KS}$. In Eq.~(\ref{eq:ks-equation}), $\hat{t}_\t{s}$ denotes the kinetic energy operator, $\hat{v}_\t{ext}$ the external potential, $\hat{v}_\t{H}$ the Hartree potential, and $\hat{v}_\t{xc}$ the exchange-correlation potential. Solving Eq.~(\ref{eq:ks-equation}) yields the Kohn-Sham single particle states~$\psi_p$ and their eigenenergies~$\epsilon_{p}$. For a spin-unpolarized system, these states determine the electron density via
\begin{equation}
n(\mathbf{r})=\sum_{p} f(\epsilon_{p}) |\psi_p(\mathbf{r})|^2,
\label{eq:density}
\end{equation}
whereby the occupation numbers~$f(\epsilon_{p})$ are chosen in such a way that the $N_e/2$ states with the lowest eigenvalues~$\epsilon_{p}$
are doubly occupied.

To solve Eq.~(\ref{eq:ks-equation}) in numerical implementations,
the Kohn-Sham states are expanded in a finite basis set~$\chi_\mu(\vec{r}-\vec{R}_{I(\mu)})$ as
\begin{equation}
\psi_p(\mathbf{r})=\sum_{\mu}C_{\mu p} \: \chi_{\mu}(\vec{r}-\vec{R}_{I(\mu)}),
\label{eq:expansion}
\end{equation}
with the expansion coefficients $C_{\mu p}$. The chosen notation highlights that in a numerical atom-centered
basis set, each basis function~$\mu$ is associated to an atom~$I(\mu)$ situated at~$\vec{R}_{I(\mu)}$. In such a basis set, Eq.~(\ref{eq:ks-equation}) becomes a generalized eigenvalue problem 
\begin{equation}
\sum_{\nu} H_{\mu\nu} C_{\nu p} = \epsilon_{p}  \sum_{\nu} S_{\mu\nu} C_{\nu p}\;.
\label{eq:HKS0_algebraic}
\end{equation}
Using the bra-ket notation~$\langle.|.\rangle$ for the inner product in Hilbert space,
$H_{\mu\nu}$ denotes the elements~$\braket{\chi_{\mu}|\hat{h}_\t{KS}|\chi_{\nu}}$ of 
the Hamiltonian matrix and $S_{\mu\nu}$ the elements~$\braket{\chi_{\mu}|\chi_{\nu}}$ of the overlap matrix.
Accordingly, the variation with respect to the density becomes a minimization with respect to the expansion 
coefficients~$C_{\nu p}$
\begin{equation}
E_\t{tot}^0 = E_\t{KS}[n_0(\vec{r})] = \min_{C_{\nu p}}\left[E_\t{KS}-\sum_{p}\epsilon_{p}(\braket{\psi_p|\psi_p} -1)\right] \;,
\label{eq:ks-variationalC}
\end{equation}
in which the eigenstates~$\psi_p$ are constrained to be orthonormal. Typically, the ground state density~$n_0(\vec{r})$ and the associated total energy~$E_\t{tot}$ are determined by solving Eq.~(\ref{eq:ks-variationalC}) iteratively, until self-consistency is achieved.

If an external electric field~$\vec{E}$ is applied to an isolated system, the KS Hamiltonian gains an additional term~$\hat{h}_\t{E}= - \vec{r} \cdot \vec{E}$. If this electric field~$\vec{E}=\left(e_x,e_y,e_z\right)$ 
is a superposition of homogeneous electrical fields with strengths~$e_\gamma$ aligned along the different cartesian axes~$\gamma$, the additional term~$\hat{h}_\t{E}$ contributes
\begin{equation}
E_\t{E}[n] = - \sum_\gamma \int e_\gamma r_\gamma \, n(\vec{r}) \, d\vec{r} 
\end{equation}
to the total energy functional in Eq.~(\ref{eq:KSTOT}). A perturbative Taylor-expansion of the total energy in the zero-field limit then gives
\begin{equation}
E_\t{tot}(\vec{E}) \approx E^0_\t{tot} + \sum_{\gamma}\mu_{\gamma}e_{\gamma} + \frac{1}{2}\sum_{\gamma,\delta}\alpha_{\gamma\delta}e_{\gamma}e_\delta + \cdots\;, 
\end{equation}
where $\eta,\gamma$ are Cartesian directions. For isolated systems, the coefficient in the linear term
\begin{equation}
\mu_{\gamma}= \left.\frac{\partial E_\t{E}[n_0]}{\partial e_\gamma}\right\vert_{\vec{E}=0} = - \int{n_0(\vec{r}) r_\gamma d\vec{r} },
\label{eq:dipole}
\end{equation}
which corresponds to the $\gamma$-component of the dipole moment of the system in its ground state, can be directly evaluated at the DFT level of theory due to the Hellmann-Feynman theorem. However, this is not possible for the
coefficient in the second-order term,~i.e.,~the polarizability
\begin{eqnarray}
\alpha_{\gamma\delta} 
& = &  \left.\frac{\partial^2 E_\t{E}[n_0]}{\partial e_\gamma\partial e_\delta}\right\vert_{\vec{E}=0} =    \left.\frac{\partial \mu_{\gamma}}{\partial e_{\delta}}\right\vert_{\vec{E}=0}  
 =  - \int{ r_\gamma  \left(\left.\frac{\partial n_0(\vec{r})}{\partial e_{\delta} }\right\vert_{\vec{E}=0}\right) d \vec{r}},
\label{eq:polar}
\end{eqnarray}
since the derivative (or response) of the ground-state density with respect to the field strength
is explicitly required. More generally, this is formalized within the $2n+1$ rule~\cite{Gonze-1989}, which states that second-order derivatives of the total energy~\cite{Mosley1993,Giannozzi1991,Ferrero2008} 
cannot be direcly calculated from the ground-state electron density or wavefunction alone, but also require the respective first-order derivatives of the electron density or wavefunction,~i.e.,~their linear response to the perturbation.
We use perturbation theory~(PT) to obtain the required derivatives. In this formalism, the response to perturbations along different Cartesian axes~$\gamma$ can be treated independently viz. subsequently, so that the short-hand notation 
\begin{equation}
M^{(1)}=\frac{d {M^{(0)}}}{d {e_{\gamma}}}
\end{equation}
used in the following for ground-state~$M^{(0)}$ and response properties~$M^{(1)}$ is always well-defined,~e.g.,
\begin{equation}
n^{(1)}=\frac{d {n^{(0)}}}{d {e_{\gamma}}} = \frac{d {n_{0}}}{d {e_{\gamma}}}  \;.
\end{equation}
In this way, we can express the linear Taylor-expansion of the Kohn-Sham Hamiltonian in the limit of vanishing field along the $\gamma$-axis as:
\begin{equation}
\hat{h}_\t{KS}(e_\gamma) \approx  \hat{h}_\t{KS}^{(0)} + \hat{h}_\t{KS}^{(1)}e_\gamma + \cdots\;,
\end{equation}
where the response of the Hamiltonian operator is 
\begin{equation}
\hat{h}_\t{KS}^{(1)} = \hat{v}_\t{ext}^{(1)}(r)+\hat{v}_\t{H}^{(1)}+\hat{v}_\t{xc}^{(1)}-r_\gamma \,.
\label{eq:H1}
\end{equation}
Introducing the analogous expansions 
\begin{equation}
\psi_p({e_\gamma}) \approx  \psi_p^{(0)} + \psi_p^{(1)}e_\gamma + \cdots \quad\quad\quad
\epsilon_{p}(e_\gamma) \approx  \epsilon_{p}^{(0)} + \epsilon_{p}^{(1)}e_\gamma + \cdots
\end{equation}
for the single-particle states~$\psi_p({e_\gamma})$ and their eigenvalues~$\epsilon_{p}(e_\gamma)$, rearranging the linear-order terms in the KS equation~$\hat{h}_\t{KS}(e_\gamma)\psi_p({e_\gamma})=\epsilon_{p}(e_\gamma)\psi_p({e_\gamma})$, and applying the normalization condition~$\braket{\psi_p({e_\gamma})|\psi_p({e_\gamma})}=1$, yields the \emph{Sternheimer equation}
\begin{equation}
  \left(\hat{h}_\t{KS}^{(0)} - \epsilon_{p}^{(0)}\right) \ket{\psi_p^{(1)}} =
- \left(\hat{h}_\t{KS}^{(1)} - \epsilon_{p}^{(1)}\right) \ket{\psi_p^{(0)}}  \;,
\label{eq:Sternheimer}
\end{equation}
as well as the condition
\begin{equation}
\braket{\psi_p^{(1)}|\psi_p^{(0)}} + \braket{\psi_p^{(0)}|\psi_p^{(1)}} = 0 \;.
\label{eq:norm_psi_1}
\end{equation}
By multiplying Eq.~(\ref{eq:Sternheimer}) with~$\bra{\psi_q^{(0)}}$ from the left one obtains
\begin{equation}
\left(\epsilon_{q}^{(0)} - \epsilon_{p}^{(0)}\right) \braket{\psi_q^{(0)}|\psi_p^{(1)}}  =  - \left( \braket{\psi_q^{(0)}|\hat{h}_\t{KS}^{(1)}|\psi_p^{(0)}} - \epsilon_{p}^{(1)} \delta_{qp}\right)  \;.
\label{eq:SternheimerPQ}
\end{equation}
To solve this equation numerically, we expand the response of the wave functions 
\begin{eqnarray}
\psi_p^{(1)}(\mathbf{r}) & = & \sum_{q} U_{qp}^{(1)}\psi_q^{(0)}(\mathbf{r}) =  \sum_{\mu} \underbrace{\sum_q U_{qp}^{(1)}C_{\mu q}^{(0)}}_{C_{\mu p}^{(1)}} \: \chi_{\mu}(\mathbf{r})
\label{eq:expansion_PBC_fst_order}
\end{eqnarray}
in terms of the unperturbed states~$\psi_q^{(0)}(\mathbf{r})$. Here, we chose $U_{pp}^{(1)}=0$ for all~$p$ to fulfill Eq.~(\ref{eq:norm_psi_1}) and hence obtain an  algebraic expression for Eq.~(\ref{eq:SternheimerPQ}) 
\begin{equation}
  \left(\epsilon_{q}^{(0)} - \epsilon_{p}^{(0)}\right)   U_{qp}^{(1)}    =  
- \sum_{\mu\nu} \left(C_{\mu q}^{(0)}\right)^* C_{\nu p}^{(0)} \braket{\chi_{\mu}|\hat{h}_\t{KS}^{(1)}|\chi_{\nu}} + \epsilon_{p}^{(1)} \delta_{qp}\;.
\end{equation}
The expansion using the matrix~$U_{qp}^{(1)}$ employed in this work is typical for the coupled perturbed self-consistent field~(CPSCF) formulation~\cite{Gerratt-1967,Pople-1979,Dykstra-1984,Frisch-1990, Ochsenfeld-1997, Liang-2005} of PT. For our implementation described in Sec.~\ref{sec:Implementation}, such an expansion in terms of orbitals is advantageous, since it allows leveraging the already existing algorithms for the massively-parallel evaluation of matrix elements in this representation~\cite{Blum2009,Havu/etal:2009,Shang:2017hn}. Accordingly, the matrix elements~$H_{\mu\nu}^{(1)}=\braket{\chi_{\mu}|\hat{h}_\t{KS}^{(1)}|\chi_{\nu}}$ are defined as for unperturbed calculations,~i.e.,~using the numeric atomic orbitals introduced in Eq.~(\ref{eq:expansion}). This allows us to directly compute the non-diagonal elements~($q\neq p$) of
\begin{equation}
 U_{qp}^{(1)}    = \frac{\sum_{\mu\nu} \left(C_{\mu q}^{(0)}\right)^* \;  H_{\mu\nu}^{(1)} \; C_{\nu p}^{(0)}}{\epsilon_{p}^{(0)} - \epsilon_{q}^{(0)}}   \;.
\label{eq:U1}
\end{equation}
The matrix~$U_{qp}^{(1)}$, which fulfills~$U_{qp}^{(1)}=-\left(U_{pq}^{(1)}\right)^*$, plays a central role in our implementation: As discussed in detail in Sec.~\ref{sec:Implementation}, it allows us to directly
determine the response of the density 
\begin{equation}
n^{(1)}=\sum_p f(\epsilon_p) \left[ \psi_p^{(1)} \psi_p^{(0)} + \psi_p^{(0)}\psi_p^{(1)} \right] 
\label{eq:density_resp}
\end{equation}
in a density-matrix formalism,~i.e.,~without explicitly computing the response of the eigenvalues~$\epsilon_p^{(1)}$, of the wave function~$\psi_p^{(1)}(\mathbf{r})$, or its coefficients~$C_{\mu p}^{(1)}$, which is computationally advantageous. Using $U_{qp}^{(1)}$, one can then  directly evaluate the polarizability tensor~$\alpha_{\gamma\delta}$ defined in Eq.~(\ref{eq:polar}) in finite, isolated systems.
Let us note that implementations of PT in plane wave codes typically do not use the expansion in terms of orbitals defined in Eq.~(\ref{eq:expansion_PBC_fst_order}) via the $U_{qp}^{(1)}$ matrix,
but rather compute the coefficients~$C_{\mu p}^{(1)}$ by directly solving Eq.~(\ref{eq:SternheimerPQ}) in the space spanned by the KS states using the density-functional perturbation theory formalism~(DFPT)~\cite{Gonze1997-1,Gonze1997-2, PutrinoParrinelloDFPT2000, Baroni-2001}. In such codes, in which thousands of orbitals,~i.e.,~plane waves, need to be considered, the DFPT approach is advantageous.

\section{Generalization to Periodic Solids}
\label{ssec:solids}
For periodic boundary conditions~(PBCs), the main physical reasoning behind the derivation of Eqs.~(\ref{eq:KSTOT})-(\ref{eq:density_resp}) still remains valid. 
However, three specific adaptations have to be made:

First, the basis set expansion introduced in Eq.~(\ref{eq:expansion}) is slightly different, as described in detail in Refs.~\cite{Blum2009,Knuth:2015kc,Shang:2017hn}: The periodic images of the nuclei~$\vec{R}_{Im}=\vec{R}_I+\vec{R}_m$ are accounted for by summing over the lattice vectors~$\vec{R}_m$,~i.e.,~over linear combinations of the unit cell lattice vectors~$\vec{a}_1,\vec{a}_2,\vec{a}_3$. Analogously, also the numeric atomic orbitals associated with such periodic images,~e.g.,~$\chi_{\mu m}(\mathbf{r}) = \chi_{\mu}(\mathbf{r}-\vec{R}_{I(\mu)}-\vec{R}_m)$ associated with the periodic image~$m$ of nucleus~$\vec{R}_{I}$, gain an additional index~$m$ that describes their relative position to the unit-cell equivalent. To account for translational symmetry and exploit Bloch's theorem, Bloch-like generalized basis functions
\begin{equation}
\varphi_\mu(\vec{k},\mathbf{r})=\sum_{m} \chi_{\mu m}(\vec{r}) \exp\left(-i\vec{k}\cdot\vec{R}_m\right)
\label{eq:Bloch_PBC}
\end{equation}
are constructed from the local atomic orbitals and then used in the basis set expansion
\begin{equation}
\psi_p^{(0)}(\vec{k},\mathbf{r})=\sum_{\mu}C_{\mu p}^{(0)}(\vec{k}) \: \varphi_{\mu}(\vec{k},\mathbf{r}) \;.
\label{eq:expansion_PBC}
\end{equation}
Accordingly, all relevant physical quantities such as the KS Hamiltonian
\begin{equation}
H_{\mu\nu}^{(0)}(\mathbf{k})=\sum_{m,n} e^{-i\mathbf{k}\left(\mathbf{R}_n-\vec{R}_m\right)}\int\limits_\t{u.c.}\chi_{\mu m}(\vec{r})\, \hat{h}_\t{KS}\, \chi_{\nu n}(\vec{r})\, d\vec{r} 
\label{eq:HKS_PBC}
\end{equation}
gain an additional dependence on the wavevector~$\vec{k}$, so that Eq.~(\ref{eq:HKS0_algebraic}) becomes
\begin{equation}
\sum_{\nu} H_{\mu\nu}^{(0)}(\vec{k}) C_{\nu p}^{(0)}(\vec{k)} = \epsilon_{p}^{(0)}(\vec{k)}  \sum_{\nu} S_{\mu\nu}(\vec{k)} C_{\nu p}^{(0)}(\vec{k)}\;.
\end{equation}
Therefore, the summations over electronic states appearing in Eqs.~(\ref{eq:KSTOT})-(\ref{eq:density_resp}) now feature an additional analytical integration over the Brillouin zone that is approximated numerically by a sum over a finite $\vec{k}$-grid with $N_k$~points. Similarly, the real-space integrals in Eqs.~(\ref{eq:KSTOT})-(\ref{eq:density_resp}) are no longer indefinite, but definite and limited to the unit cell~(u.c.), as it is the case in Eq.~(\ref{eq:HKS_PBC}).

%
Second, it is necessary to consider the \emph{screened} electric field~$\vec{E} = \vec{D} - 4\pi \vec{P}$ in the solid,
where $\vec{D}$ is the electric displacement~\cite{Griffiths:1999ud} and the polarization in the unit cell volume~$V$ is given by~\cite{Gonze1997-1,Baroni-2001}:
\begin{equation}
{P}_\gamma =  - \frac{1}{V}\int\limits_{\t{u.c.}} r_\gamma  n_0(\vec{r})d \vec{r}\; .
\end{equation}
The relationship between the components of the electric displacement and the screened field defines the high-frequency dielectric constant~\cite{Griffiths:1999ud} 
\begin{equation}
\varepsilon_{\gamma\delta}^{\infty} =  \frac{\partial D_{\gamma}}{\partial E_\delta} = \delta_{\gamma\delta} + 4\pi \frac{\partial P_\gamma}{\partial E_\delta} \;,
\label{eq:dielec}
\end{equation}
where $\eta,\gamma$ are Cartesian directions and $\delta$ is the Kronecker delta symbol. 
For a screened field~$\vec{E}=\left(e_x,e_y,e_z\right)$ that consists of a superposition of homogeneous electrical fields with strengths~$e_\gamma$ 
aligned along the different cartesian axes~$\gamma$, one can follow the derivation given in the previous section to obtain
\begin{eqnarray}
\left.\frac{\partial E_\t{E}[n_0]}{\partial e_\gamma}\right\vert_{\vec{E}=0}                    & = &  -\int\limits_{\t{u.c.}}  r_\gamma n_0(\vec{r}) d\vec{r}  = VP_\gamma \rightarrow \mu_\gamma  \\
\left.\frac{\partial^2 E_\t{E}[n_0]}{\partial e_\gamma\partial e_\delta}\right\vert_{\vec{E}=0} & = &  -\int\limits_{\t{u.c.}}{ r_\gamma  \left(\left.\frac{\partial n_0(\vec{r})}{\partial e_{\delta} }\right\vert_{\vec{E}=0}\right) d \vec{r}} =  V\left.\frac{\partial P_{\gamma}}{\partial e_{\delta}}\right\vert_{\vec{E}=0} \rightarrow \alpha_{\gamma\delta} \; .
\label{derPolarization}
\end{eqnarray}
A comparison with Eqs.~(\ref{eq:dipole}) and~(\ref{eq:polar}) reveals the formal relationship between the dipoles~$\mu_\gamma$ and the polarizabilities~$\alpha_{\gamma\delta}$ discussed
in the previous section for molecules and the polarization~$P_\gamma$,~i.e.,~a dipole density~\cite{ashcroft}, and its derivative with respect to the screened field in solids.
For the sake of notational clarity, the ``molecular'' notation with $\mu_\gamma$ and $\alpha_{\gamma\delta}$ is used for the remainder of this paper.

Third, complications arise due to the fact that the superposition of homogeneous electric fields~$\vec{E}$ is not periodic, as alluded to in the introduction. As a consequence, the definite integral over the unit cell required to determine the Hamiltonian response~$H_{\mu\nu}^{(1)}(\vec{k})$ is ill-defined in PBCs, since
$\hat{h}_\t{KS}^{(1)}$ given in Eq.~(\ref{eq:H1}) contains the position operator~${r_\gamma}$, which is itself ill-defined in this case. The same problem affects Eq.~(\ref{derPolarization}). 
In reciprocal space implementations, the Berry-phase formalism~\cite{Vanderbilt1993,Resta1994,PutrinoParrinelloDFPT2000,Mosley1993,Gonze1997-2,Ferrero2008} is typically the method of choice; a tutorial introduction to this approach can be found in Ref.~\cite{Spaldin:2012ez}. In real space implementations, the position operator can be rewritten in a boundary-insensitive form~\cite{Giannozzi1991} by exploiting the properties of the commutator between the KS-Hamiltonian and the position operator~$\left[\hat{h}_\t{KS}^{(0)}(\vec{k}),\vec{r}\right]=-\nabla$. With that, one gets the well-known expression
\begin{eqnarray}
\braket{\psi_{q}^{(0)}(\mathbf{k}) \vert \nabla_\gamma \vert \psi_{p}^{(0)}(\mathbf{k})} 
 & = &  - \braket{\psi_{q}^{(0)}(\mathbf{k}) \vert \left[ \hat{h}_\t{KS}^{(0)}(\vec{k}),{r_\gamma} \right] \vert \psi_{p}^{(0)}(\mathbf{k})} \\
 & = &  \left( \epsilon_p^{(0)}(\mathbf{k})-\epsilon_q^{(0)}(\mathbf{k}) \right) \braket{\psi_{q}^{(0)}(\mathbf{k}) \vert {r_\gamma} \vert \psi_{p}^{(0)}(\mathbf{k})} \;,
\end{eqnarray}
that can be used to evaluate the non-diagonal matrix elements~($q \neq p$)
\begin{equation}
\braket{\psi_{q}^{(0)}(\mathbf{k}) \vert {r_\gamma} \vert \psi_{p}^{(0)}(\mathbf{k})}  =  \frac{\braket{\psi_{q}^{(0)}(\mathbf{k}) \vert \nabla_\gamma \vert \psi_{p}^{(0)}(\mathbf{k})}}{ \epsilon_p^{(0)}(\mathbf{k})-\epsilon_q^{(0)}(\mathbf{k})} \;.
\end{equation}
Using Eqs.~(\ref{eq:Bloch_PBC}) and~(\ref{eq:expansion_PBC}) we obtain the representation
\begin{equation}
\Omega_{qp}(\vec{k})=
-\braket{\psi_{q}^{(0)}(\mathbf{k}) \vert {r_\gamma} \vert \psi_{p}^{(0)}(\mathbf{k})}  = -\sum_{\mu\nu} \frac{\left(C_{\mu q}^{(0)}(\vec{k})\right)^*C_{\nu p}^{(0)}(\vec{k}) }{\epsilon_p^{(0)}(\mathbf{k})-\epsilon_q^{(0)}(\mathbf{k}) } R_{\mu\nu}^{(0)}(\vec{k})
\label{eq:wPBC}
\end{equation}
with
\begin{equation}
R_{\mu\nu}^{(0)}(\vec{k})=\sum_{mn}e^{-i\mathbf{k}\left(\mathbf{R}_n-\vec{R}_m\right)}\int\limits_\t{u.c.}\chi_{\mu m}(\vec{r})\, \nabla_\gamma \, \chi_{\nu n}(\vec{r})\, d\vec{r} \;. \label{eq:rPBC}
\end{equation}
With that we can recast the expectation value~$H_{qp}^{(1)}(\vec{k})=\braket{\psi_q^{(0)}(\vec{k})|\hat{h}_\t{KS}^{(1)}|\psi_p^{(0)}(\vec{k})}$ appearing in Eq.~(\ref{eq:SternheimerPQ}):
\begin{eqnarray}
\fl\braket{\psi_q^{(0)}(\vec{k})|\hat{h}_\t{KS}^{(1)}|\psi_p^{(0)}(\vec{k})}  & = & \braket{\psi_q^{(0)}(\vec{k})| \hat{v}_\t{ext}^{(1)}(r)+\hat{v}_\t{H}^{(1)}+\hat{v}_\t{xc}^{(1)} |\psi_p^{(0)}(\vec{k})} -  \braket{\psi_q^{(0)}(\vec{k})|r_\gamma|\psi_p^{(0)}(\vec{k})} \\
  & = &  \sum_{\mu\nu}\left(C_{\mu q}^{(0)}(\vec{k})\right)^*C_{\nu p}^{(0)}(\vec{k}) V_{\mu\nu}^{(1)}(\vec{k}) +\Omega_{qp}(\vec{k}) \;.
\label{eq:splitV_R}
\end{eqnarray}
Here, the matrix elements~$V_{\mu\nu}^{(1)}(\vec{k})=\braket{\varphi_{\mu}(\vec{k})| \hat{v}_\t{ext}^{(1)}(r)+\hat{v}_\t{H}^{(1)}+\hat{v}_\t{xc}^{(1)} |\varphi_{\nu}(\vec{k})}$ can be directly evaluated as done in Eq.~(\ref{eq:HKS_PBC}) 
\begin{equation}
V_{\mu\nu}^{(1)}(\mathbf{k})=\sum_{m,n} e^{-i\mathbf{k}\left(\mathbf{R}_n-\vec{R}_m\right)}\int\limits_\t{u.c.}\chi_{\mu m}(\vec{r})\, \left( \hat{v}_\t{ext}^{(1)}(r)+\hat{v}_\t{H}^{(1)}+\hat{v}_\t{xc}^{(1)} \right) \, \chi_{\nu n}(\vec{r})\, d\vec{r} \;,
\label{eq:V1}
\end{equation}
since they only depend on lattice periodic operators. Now, the matrix~$U_{qp}^{(1)}(\vec{k})$ introduced in Eq.~(\ref{eq:U1}) is computed as 
\begin{equation}
 U_{qp}^{(1)}(\vec{k})    = \frac{\sum_{\mu\nu} \left(C_{\mu q}^{(0)}(\vec{k})\right)^* \;  V_{\mu\nu}^{(1)}(\vec{k}) \; C_{\nu p}^{(0)}(\vec{k})}{\epsilon_{p}^{(0)}(\vec{k}) - \epsilon_{q}^{(0)}(\vec{k})}   + \frac{\Omega_{qp}(\vec{k})}{\epsilon_{p}^{(0)}(\vec{k}) - \epsilon_{q}^{(0)}(\vec{k})}  \;.
\label{eq:cpscf-LCAO}
\end{equation}

Similarly, the polarizability tensor components appearing in Eq.~(\ref{eq:polar}) can be rewritten as
\begin{eqnarray}
\alpha_{\gamma\delta} & = &  -\int\limits_{\t{u.c.}}{ r_\gamma  \left(\left.\frac{\partial n_0(\vec{r})}{\partial e_{\delta} }\right\vert_{\vec{E}=0}\right) d \vec{r}} \nonumber\\ 
& = & -\frac{1}{N_k}\sum_{p,\vec{k}}   f(\epsilon_p(\vec{k})) \left[ \braket{\psi_p^{(1,\delta)}(\vec{k})|r_\gamma|\psi_p^{(0)}(\vec{k})} + \braket{\psi_p^{(0)}(\vec{k})|r_\gamma|\psi_p^{(1,\delta)}(\vec{k})} \right] \nonumber\\
& = &  \frac{1}{N_k}\sum_{q,p,\vec{k}} f(\epsilon_p(\vec{k})) 
\left[ 
\left(U_{qp}^{(1,\delta)}(\vec{k})\right)^*\Omega_{qp}^{(\gamma)}(\vec{k})\; + \;U_{qp}^{(1,\delta)}(\vec{k}) \Omega_{pq}^{(\gamma)}(\vec{k}) 
\right]
\label{eq:polarizability_U}
\end{eqnarray}
using the matrix elements defined in Eq.~(\ref{eq:wPBC}). As explicitly highlighted in the notation, the matrix~$U_{qp}^{(1,\delta)}(\vec{k})$ associated with a perturbation along the Cartesian axis~$\delta$ has to be used in this case, whereas the matrix~$\Omega_{qp}^{(\gamma)}(\vec{k})$ is associated with a perturbation along the Cartesian axis~$\gamma$. Throughout the remainder of this work, the more general formulation in terms of Bloch-functions~$\varphi_\mu(\vec{k})$ and wave vectors~$\vec{k}$ is used, since a simplification to finite systems is straightforward.

\section{Details of the Implementation}
\label{sec:Implementation}

\begin{figure}[t]
\centering
\includegraphics[width=0.50\columnwidth]{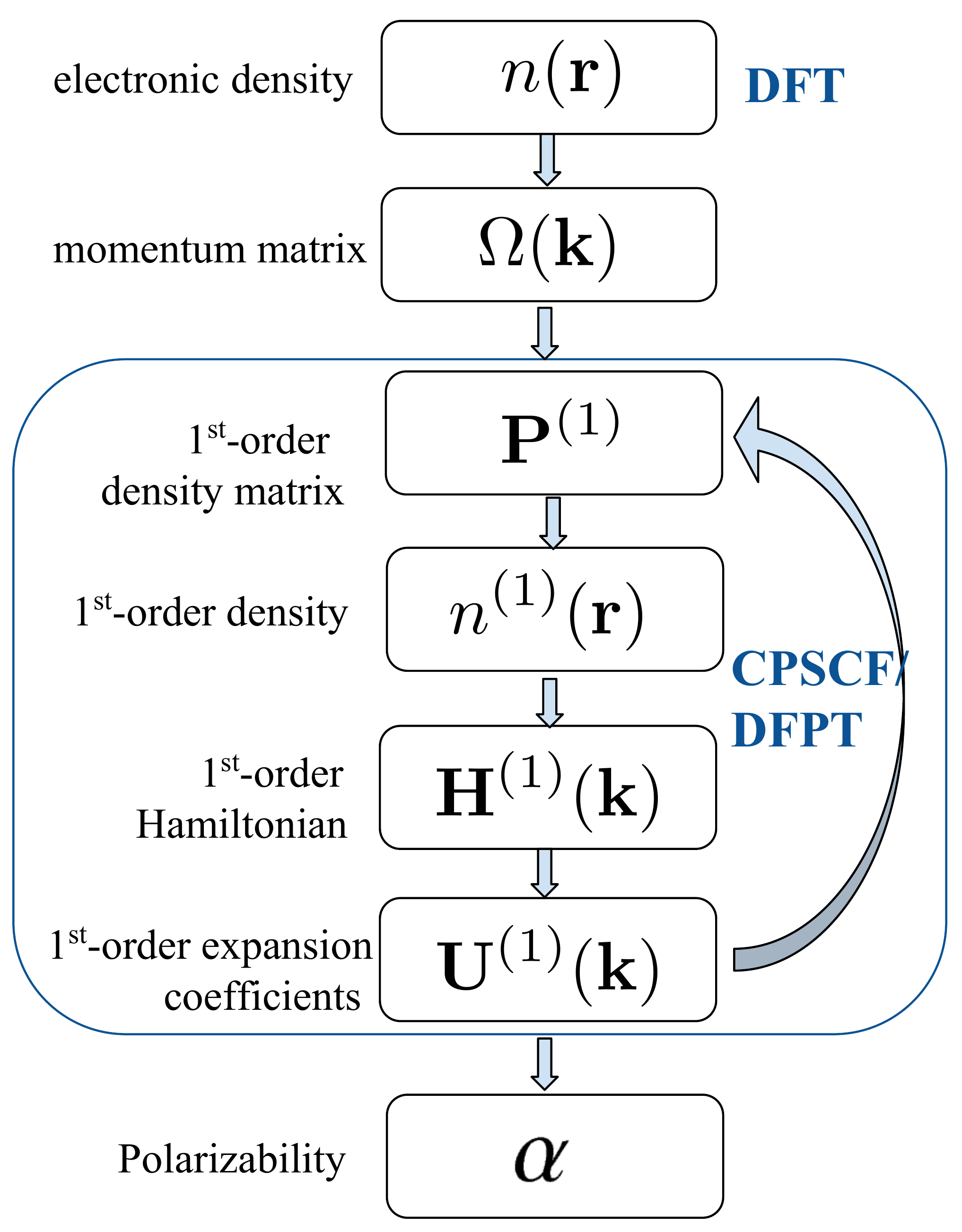}
\caption{Flowchart for the calculation of the polarizability. Loops are performed over the different Cartesian coordinates and, in the case of periodic boundary
conditions, over the finite $\vec{k}$-grid.}
\label{fig:DFPT_flowchart_E_filed}
\end{figure}

Our implementation closely follows the flowchart shown in Fig.~\ref{fig:DFPT_flowchart_E_filed}: 
After a ground state DFT calculation (see Ref.~\cite{Blum2009}) is completed, the matrix~$\Omega_{qp}(\vec{k})$ 
is computed. If $U_{pq}^{(1)}(\vec{k})=0$ is used as initial guess, one obtains
\begin{equation}
 U_{qp}^{(1)}(\vec{k})    =  \frac{\Omega_{qp}(\vec{k})}{\epsilon_{p}^{(0)}(\vec{k}) - \epsilon_{q}^{(0)}(\vec{k})}  
\end{equation}
in the first iteration, which can then be fed back to the self-consistency loop to determine the first-order density response~$n^{(1)}(\vec{r})$
in a density matrix formalism~(see Sec.~\ref{sec:n1}). As detailed in Sec.~\ref{ssec:hKS1}, we then use~$n^{(1)}(\vec{r})$ to compute the remaining, individual ingredients that enter~$\braket{\psi_q^{(0)}(\vec{k})|\hat{h}_\t{KS}^{(1)}|\psi_p^{(0)}(\vec{k})}$,~i.e.,~the matrix elements~$V_{\mu\nu}^{(1)}(\vec{k})$ defined in Eq.~(\ref{eq:V1}). The Sternheimer equation  then provides a new matrix~$U_{qp}^{(1)}(\vec{k})$, as discussed in Sec.~\ref{sec:C1}. We iteratively restart the PT loop using a Pulay-mixer~\cite{Pulay1980}, until self-consistency is reached,~i.e.,~until the changes in the response of the density-matrix~$\vec{P}^{(1)}$ become smaller than a user-given threshold. In the last step, the polarizability and the dielectric constant are computed, as discussed in Sec.~\ref{ssec:alpha}. Atomic units are used in the complete workflow.

\begin{figure}
\centering
\includegraphics[width=0.95\columnwidth]{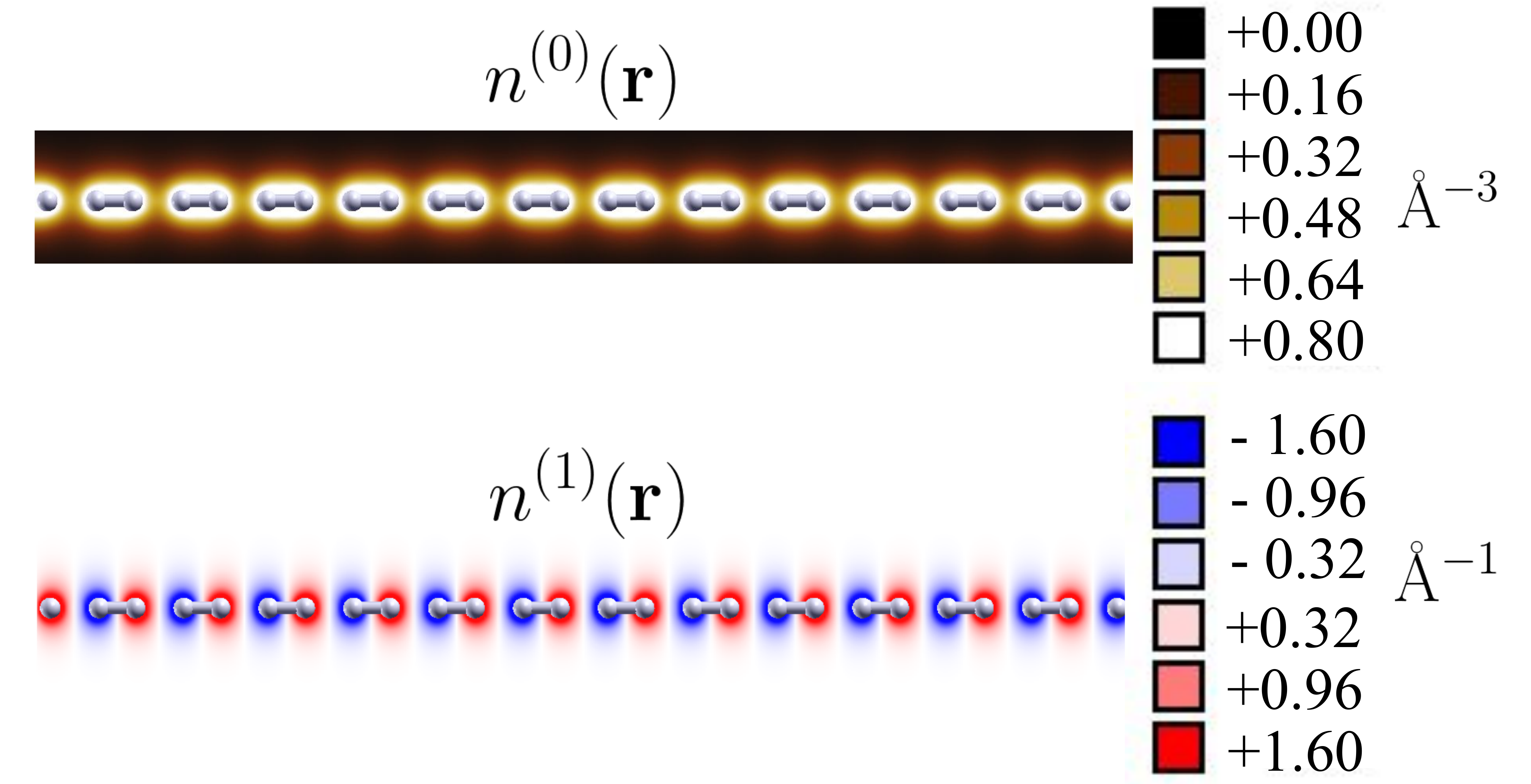}
\caption{Ground-state electronic density~$n^{(0)}(\vec{r})$ and its response~$n^{(1)}(\vec{r})$ to an electric field, as exemplarily computed for an infinite, periodic H$_2$ chain.}
\label{fig:H2_line_rho1}
\end{figure}

Both the ground-state density~$n^{(0)}(\mathbf{r})$ and the
response of the density~$n^{(1)}(\mathbf{r})$ are periodic,~i.e.,~invariant against translations
\begin{equation}
n^{(0)}(\mathbf{r}+\vec{R}_m)=n^{(0)}(\mathbf{r}) \quad \quad n^{(1)}(\mathbf{r}+\vec{R}_m)= n^{(1)}(\mathbf{r}) 
\end{equation}
by a lattice vector~$\vec{R}_m$, as shown in Fig.~\ref{fig:H2_line_rho1}. 
Accordingly, we can use the algorithms used in ground-state calculations and discussed in detail in Refs.~\cite{Blum2009,Knuth:2015kc} for many aspects of our implementation. In the following, we thus mainly focus on the practical details that are specifically needed for the computation of the response to a homogeneous electric field.

\subsection{Response of the Electronic Density}
\label{sec:n1}

To numerically compute the electronic density~$n(\vec{r})$ in ground-state calculations~\cite{Blum2009}, we use a density matrix formalism
\begin{equation}
n^{(0)}(\mathbf{r})= \sum_{\mu m,\nu n}{P_{\mu m,\nu n}^{(0)}\chi_{\mu m}^{(0)}(\mathbf{r})\chi_{\nu n}^{(0)}}(\mathbf{r}) \;,
\end{equation}
which is obtained by inserting Eqs.~(\ref{eq:expansion_PBC}) and~(\ref{eq:Bloch_PBC}) into Eq.~(\ref{eq:density}). Hence, the density matrix is given by
\begin{eqnarray}
\fl P_{\mu m,\nu n}^{(0)} & = & \frac{1}{N_k}\sum_{\mathbf{k}}\left[ e^{-i\mathbf{k}(\vec{R}_n-\mathbf{R}_m)} \sum_{o} f(\epsilon_o(\vec{k})) \left(C_{\mu o}^{(0)}(\mathbf{k})\right)^* C_{\nu o}^{(0)}(\mathbf{k}) \right]   \;.
\end{eqnarray}
Here, the chosen notation using the index~$o$ highlights that the sum over \textbf{all} states only needs to be performed over occupied states with~$f(\epsilon_o(\vec{k}))\neq 0$ in practice.
Similarly, the response of the electronic density can thus be expressed as 
\begin{eqnarray}
n^{(1)}(\mathbf{r}) & = & \sum_{\mu m,\nu n}{P_{\mu m,\nu n}^{(1)}\chi_{\mu m}^{(0)}(\mathbf{r})\chi_{\nu n}^{(0)}}(\mathbf{r}) 
\label{eq:n1}
\end{eqnarray}
using the response of the density matrix given by
\begin{eqnarray}
\fl P_{\mu m,\nu n}^{(1)}  =   \frac{1}{N_k}\sum_{\mathbf{k}}\left\{ e^{-i\mathbf{k}(\vec{R}_n-\mathbf{R}_m)} \sum_{o} f(\epsilon_o(\vec{k})) \left[  \left(C_{\mu o}^{(1)}(\mathbf{k})\right)^* C_{\nu o}^{(0)}(\mathbf{k}) \right.\right. \nonumber\\
 + \left.\left. \left(C_{\mu o}^{(0)}(\mathbf{k})\right)^* C_{\nu o}^{(1)}(\mathbf{k}) \right] \right\}  \;.
\label{CChere}
\end{eqnarray}
In the sum over states~$o$, we express $\vec{C}^{(1)}(\vec{k})$ in terms of~$\vec{U}^{(1)}(\vec{k})$ via
\begin{eqnarray}
\fl
\label{eq:expansion_PBC_fst_order_uo}
C_{\mu p}^{(1)}(\vec{k}) & = & \sum_{q} U_{qp}^{(1)}(\vec{k})C_{\mu q}^{(0)}(\vec{k}) \\ 
 & = & \frac{1}{2}\sum_{o'} f(\epsilon_{o'}(\vec{k})) U_{o'p}^{(1)}(\vec{k})C_{\mu o'}^{(0)}(\vec{k}) + \frac{1}{2}\sum_u \left(2-f(\epsilon_u(\vec{k}))\right) U_{up}^{(1)}(\vec{k})C_{\mu u}^{(0)}(\vec{k}) \;,\nonumber
\end{eqnarray}
whereby we split the sum over~$q$ into two separate sums over~$o'$ and~$u$. In practice, these two sums can then be later evaluated by restricting the sum over~$o'$ to occupied and the sum over~$u$ to unoccupied states, respectively.
Accordingly, also the sum over~$o$ appearing in Eq.~(\ref{CChere}) can be split into two double sums, one over~$o,o'$ and one over~$o,u$. For the first one, we obtain 
\begin{eqnarray}
\label{eq:oo_zero}
\fl
\sum_{o,o'} \frac{f(\epsilon_{o}(\vec{k}))f(\epsilon_{o'}(\vec{k}))}{2} \left[\left(C_{\mu o'}^{(0)}(\mathbf{k})  U_{o'o}^{(1)}(\mathbf{k}) \right)^* C_{\nu o}^{(0)}(\mathbf{k}) + \left(C_{\mu o}^{(0)}(\mathbf{k})\right)^* U_{o'o}^{(1)}(\mathbf{k}) C_{\nu o'}^{(0)}(\mathbf{k}) \right] & = & \\
\fl
\sum_{o,o'} \frac{f(\epsilon_{o}(\vec{k}))f(\epsilon_{o'}(\vec{k}))}{2} \left[\left(C_{\mu o'}^{(0)}(\mathbf{k})  U_{o'o}^{(1)}(\mathbf{k}) \right)^* C_{\nu o}^{(0)}(\mathbf{k}) + \left(C_{\mu o'}^{(0)}(\mathbf{k})\right)^* U_{oo'}^{(1)}(\mathbf{k}) C_{\nu o}^{(0)}(\mathbf{k}) \right] & = &  0 \;, \nonumber
\end{eqnarray}
which vanishes due to~$U_{oo'}^{(1)}(\vec{k})=-\left(U_{o'o}^{(1)}(\vec{k})\right)^*$, cf.~Eq.~(\ref{eq:cpscf-LCAO}). 
For the second double sum, we obtain
\begin{eqnarray}
\fl
\sum_{o,u} \frac{f(\epsilon_o(\vec{k}))\left(2-f(\epsilon_u(\vec{k}))\right)}{2}  \left[\left(C_{\mu u}^{(0)}(\mathbf{k})  U_{uo}^{(1)}(\mathbf{k}) \right)^* C_{\nu o}^{(0)}(\mathbf{k}) + \left(C_{\mu o}^{(0)}(\mathbf{k})\right)^* U_{uo}^{(1)}(\mathbf{k}) C_{\nu u}^{(0)}(\mathbf{k})  \right] \\ 
= \sum_{o,u} \left( f(\epsilon_o(\vec{k}))-f(\epsilon_u(\vec{k}))\right) \left(C_{\mu u}^{(0)}(\mathbf{k})  U_{uo}^{(1)}(\mathbf{k}) \right)^* C_{\nu o}^{(0)}(\mathbf{k}) \; 
\end{eqnarray}
by switching the summation indices~$u,o$ in the second term and using~$U_{ou}^{(1)}(\vec{k})=-\left(U_{uo}^{(1)}(\vec{k})\right)^*$, as done already for Eq.~(\ref{eq:oo_zero}).
By this means, the response of the density matrix can be written as 
\begin{eqnarray}
\fl P_{\mu m,\nu n}^{(1)}  =   \frac{1}{N_k}\sum_{\mathbf{k}} e^{-i\mathbf{k}(\vec{R}_n-\mathbf{R}_m)} \sum_{o,u} (f(\epsilon_o(\vec{k})) - f(\epsilon_u(\vec{k}))) \left[ \left(C_{\mu u}^{(0)}(\mathbf{k})  U_{uo}^{(1)}(\mathbf{k}) \right)^* C_{\nu o}^{(0)}(\mathbf{k}) \right].
\label{eq:p-with-occ}
\end{eqnarray}
In practice, the evaluation of Eq.~(\ref{eq:p-with-occ}) can thus be restricted to the double sum over occupied~$o$ and unoccupied states~$u$.

\subsection{Response of the Kohn-Sham Hamiltonian}
\label{ssec:hKS1}
As discussed in Sec.~\ref{ssec:solids} for Eq.~(\ref{eq:splitV_R}), the computation of $\braket{\psi_q^{(0)}(\vec{k})|\hat{h}_\t{KS}^{(1)}|\psi_p^{(0)}(\vec{k})}$ is split 
into different steps: The matrices~$R_{\mu\nu}^{(0)}(\vec{k})$ and~$\Omega_{\mu\nu}^{(0)}(\vec{k})$, which are defined in Eqs.~(\ref{eq:wPBC})-(\ref{eq:rPBC}) and which are required to calculate~$\braket{\psi_q^{(0)}(\vec{k})|-r_\gamma|\psi_p^{(0)}(\vec{k})}$, are computed before the self-consistency loop, since they only depend on unperturbed properties. The definite unit-cell integral appearing in Eq.~(\ref{eq:rPBC}) is integrated on a real-space grid using the formalisms described in Refs.~\cite{Blum2009,Havu/etal:2009}. Conversely, the matrix~$V_{\mu\nu}^{(1)}(\vec{k})$, which is defined in Eq.~(\ref{eq:V1}) and which is required to compute $\braket{\psi_q^{(0)}(\vec{k})|\hat{v}_\t{ext}^{(1)}(r)+\hat{v}_\t{H}^{(1)}+\hat{v}_\t{xc}^{(1)}|\psi_p^{(0)}(\vec{k})}$, explicitly depends on the response of the density~$n^{(1)}(\vec{r})$ and thus needs to be updated each cycle. For that purpose, we first compute its ingredients on a real-space grid,~i.e.,~the response of the electrostatic potentials~$\hat{v}_\t{ext}^{(1)}(\vec{r})$ and $\hat{v}_\t{H}^{(1)}(\vec{r})$ as well as the response of the exchange-correlation potential~$\hat{v}_\t{xc}^{(1)}(\vec{r})$, as discussed below. The matrix elements~$V_{\mu\nu}^{(1)}(\vec{k})$ are then again obtained by performing the real-space unit-cell integral appearing in Eq.~(\ref{eq:V1}) with the aforementioned techniques.

\subsubsection{Response of the Electrostatic Potentials}
\label{sec:Ves1} 

As discussed in detail in Refs.~\cite{Blum2009,Knuth:2015kc,Shang:2017hn}, the electrostatic potential generated by the nuclei and the electrons is computed in FHI-aims ground-state calculations using a  scheme proposed by Delley~\cite{Delley1990}: The ground-state density~$n^{(0)}(\vec{r})$ is decomposed into two terms
\begin{equation}
n^{(0)}(\vec{r})= \sum_{Im} n_{Im}^{\t{free}}(\vec{r}-\vec{R}_{Im}) + \delta n(\vec{r}) \;.
\end{equation} 
The first term describes the density associated with a superposition of ``free'',~i.e.,~completely isolated, spherically symmetric atoms~$n_{Im}^{\t{free}}(\vec{r})$ located at the positions of the nuclei and of their periodic images~$\vec{R}_{Im}$. The potentials of~$n_{Im}^{\t{free}}(\vec{r})$ and $\delta  n(\vec{r})$ are computed independently and then reassembled to get the full electrostatic potential that enters the Kohn-Sham equations. For this purpose, $\delta n(\vec{r})$ is further decomposed into atom-specific multipoles, the contributions of which are added up in an Ewald-like summation to account for long-range interactions, cf. Refs.~\cite{Blum2009,Knuth:2015kc,Delley1990}. Given that the density response~$n^{(1)}(\vec{r})$ is also periodic in the perturbed case, see Fig.~\ref{fig:H2_line_rho1}, we can use the exact same formalism to obtain the electrostatic potential associated with it. There is only one small difference: In this case, the ``free'', spherically symmetric atoms do not contribute to the associated electrostatic potential at all.

\subsubsection{Response of the Exchange-Correlation Potential}

In semi-local approximations, the exchange-correlation potential~$\hat{v}_\t{xc}(\vec{r})$ entering the Kohn-Sham Hamiltonian in Eq.~(\ref{eq:ks-equation}) is given by 
\begin{equation}
  \hat{v}_\t{xc}(\vec{r})=\frac{\partial E_\t{xc}[n(\vec{r})] }{\partial n(\vec{r})} \;.
\end{equation}
Accordingly, its response~$\hat{v}_\t{xc}^{(1)}(\vec{r})$ can be obtained via
\begin{eqnarray}
\hat{v}_\t{xc}^{(1)}(\vec{r}) & = & \int \! \!  d\vec{r'} \frac{\partial^2 E_\t{xc} }{\partial  n(\vec{r}) \partial n(\vec{r'})} \frac{\partial n(\mathbf{r'})}{\partial e_\gamma} =  \int  \! \! d\vec{r'} f_\t{xc}(\vec{r},\vec{r'}) \: n^{(1)}(\mathbf{r}') \; .
\end{eqnarray}
by integrating over the the exchange-correlation kernel $f_\t{xc}(\vec{r},\vec{r}')$,~i.e.,~the second functional derivative of the exchange-correlation energy~$E_\t{xc}[n(\vec{r})]$, and the density response~$n^{(1)}(\mathbf{r}')$. For the local-density approximation~(LDA)~\cite{Perdew/Zunger:1981,Ceperley/Alder:1980} and the PBE functional~\cite{Perdew-1,Perdew-2} in the generalized-gradient approximation~(GGA), we have implemented the standard expressions for $f_\t{xc}(\vec{r},\vec{r}')$. Additionally, many more exchange-correlation kernels are accessible in our implementation via the \textit{Libxc} library~\cite{Marques:2012bu}.  

For isolated systems, we have also implemented the response of the exact-exchange potential. For Hartree-Fock and hybrid functionals, an additional exchange term  
\begin{equation}
  \left[{V}_{\t{HFX}}^{(1)}\right]_{\mu,\nu}=-\frac{1}{2}\sum_{\lambda,\sigma}P^{(1)}_{\lambda,\sigma }(\chi_{\mu}\chi_{\lambda}|\chi_{\nu}\chi_{\sigma})\; 
\end{equation}
needs to be added to the entries~$H_{\mu,\nu}^{(1)}$ of the Hamiltonian response matrix. Here, 
$(\chi_{\mu}\chi_{\lambda}|\chi_{\nu}\chi_{\sigma})$ is the two-electron, four-index Coulomb integral defined and discussed in Refs.~\cite{Ren/etal:2012,Levchenko2015,Johnson1994} and $\vec{P}^{(1)}$ is the first order density matrix defined in Eq.~(\ref{eq:p-with-occ}).

\subsection{Stable Evaluation of the Expansion Matrix~$\vec{U}^{(1)}(\vec{k})$}
\label{sec:C1}
To compute~$\vec{U}^{(1)}(\vec{k})$, one can in principle just evaluate Eq.~(\ref{eq:cpscf-LCAO}) as discussed in the beginning of Sec.~\ref{sec:Implementation}. Thereby, only the entries 
\begin{equation}
U_{uo}^{(1)}(\vec{k})    = \frac{1}{\epsilon_{o}^{(0)}(\vec{k}) - \epsilon_{u}^{(0)}(\vec{k})}\left[ \sum_{\mu\nu} \left(C_{\mu u}^{(0)}(\vec{k})\right)^* \;  V_{\mu\nu}^{(1)}(\vec{k}) \; C_{\nu o}^{(0)}(\vec{k})  + \Omega_{uo}(\vec{k}) \right] 
\label{eq:cpscf-LCAO-uo}
\end{equation}
associated to unoccupied-occupied~($uo$) orbital pairs need to be computed, since these are the only entries that enter the response of the density matrix~$\vec{P}^{(1)}$, as shown and discussed for Eq.~(\ref{eq:p-with-occ}). Obviously, Eq.~(\ref{eq:cpscf-LCAO-uo}) becomes numerically unstable when quasi-degenerate eigenvalues are present close to the Fermi energy~$\epsilon_\t{F}$, since the denominator~$\epsilon_{o}^{(0)}(\vec{k}) - \epsilon_{u}^{(0)}(\vec{k})$ approaches zero in that case. In order to overcome this difficulty, we employ the technique originally proposed by de Gironcoli~\cite{Gironcoli_PRB_1995,Baroni-2001} for DFPT-based lattice dynamics
calculations in metals. For this purpose, we use a Fermi function with a small smearing width~$\sigma$ 
\begin{equation}
\tilde{\theta}(\epsilon)=\frac{1}{1+e^{\epsilon/\sigma}} =
\frac{1}{2}\left[1-\tanh\left(\frac{\epsilon}{2\sigma}\right)\right] \;.
\end{equation} 
for the occupation numbers~$f(\epsilon_o)$ and~$f(\epsilon_u)$ appearing as difference in Eq.~(\ref{eq:p-with-occ}).
We then pull this difference~$f(\epsilon_o) - f(\epsilon_u)$ inside the evaluation of $U_{uo}^{(1)}$ and re-write the problematic prefactor in Eq.~(\ref{eq:cpscf-LCAO-uo}) as 
\begin{equation}
\fl\quad
\frac{f(\epsilon_o) - f(\epsilon_u)}{\epsilon_{o}^{(0)}(\vec{k}) - \epsilon_{u}^{(0)}(\vec{k})} \quad \longrightarrow \quad 2 \frac{
  \tilde{\theta}\left(\epsilon_\t{F}-\epsilon_{u}(\vec{k})\right)
 -\tilde{\theta}\left(\epsilon_\t{F}-\epsilon_{o}(\vec{k})\right)}{\epsilon_{o}(\vec{k})-\epsilon_{u}(\vec{k})} 
  \tilde{\theta}\left(\epsilon_{o}(\vec{k})-\epsilon_{u}(\vec{k})\right) \;,
\label{eq:U1_metal_new}
\end{equation}
as detailed in Ref.~\cite{Gironcoli_PRB_1995}. This has virtually no influence in the regime~$\epsilon_o(\vec{k})-\epsilon_{u}(\vec{k}) > \sigma$. 
For $\epsilon_{o}(\vec{k})-\epsilon_{u}(\vec{k}) \ll \sigma$, we replace and evaluate the rewritten problematic factor by its analytic limit for $\epsilon_u \rightarrow \epsilon_o$:
\begin{equation}
\fl\quad
\frac{ \tilde{\theta}\left(\epsilon_\t{F}-\epsilon_{u}(\vec{k})\right) - \tilde{\theta} \left(\epsilon_\t{F}-\epsilon_{o}(\vec{k})\right)}{\epsilon_{o}(\vec{k})-\epsilon_{u}(\vec{k})} \tilde{\theta}\left(\epsilon_{o}(\vec{k})-\epsilon_{u}(\vec{k})\right) 
\rightarrow - \frac{1}{2\sigma\left[ 1+\cosh\left(\frac{\epsilon_\textrm{\tiny F}-\epsilon_{o}(\mathbf{k})}{\sigma}\right) \right] } \;.
\label{eq:U1_metal_new_deg}
\end{equation}
This expression is always finite and therefore numerically stable, even in the case of vanishingly small energy differences. 


\subsection{Evaluation of Polarizabilities}
\label{ssec:alpha}
In the last step, we evaluate the polarizability by rewriting Eq.~(\ref{eq:polarizability_U}):
\begin{eqnarray}
\alpha_{\gamma\delta} 
& = & \frac{2}{N_k}\sum_{q,o,\vec{k}} f(\epsilon_o(\vec{k})) \;\textrm{Re}\left\{ U_{qo}^{(1,\delta)}(\vec{k}) \Omega_{oq}^{(\gamma)}(\vec{k}) \right\} \\
& = & \frac{2}{N_k}\sum_{u,o,\vec{k}} \frac{f(\epsilon_o(\vec{k}))\left(2-f(\epsilon_u(\vec{k}))\right)}{2} \;\textrm{Re}\left\{ U_{uo}^{(1,\delta)}(\vec{k}) \Omega_{ou}^{(\gamma)}(\vec{k}) \right\} \\
& = & \frac{2}{N_k}\sum_{u,o,\vec{k}} \left(f(\epsilon_o(\vec{k}))-f(\epsilon_u(\vec{k}))\right) \;\textrm{Re}\left\{ U_{uo}^{(1,\delta)}(\vec{k}) \Omega_{ou}^{(\gamma)}(\vec{k}) \right\} \,.
\label{eq:polar_bis}
\end{eqnarray}
In the first step, the use of $\Omega_{qp}(\vec{k})=\Omega_{pq}^*(\vec{k})$ reduces the summands to a real part extraction~$\textrm{Re()}$, while in the second step the same procedure as used to obtain Eq.~(\ref{eq:oo_zero}) is applied.
By this means, the double sum can be limited in practice to only run over unoccupied~$u$ and occupied~$o$ states. The same strategy introduced with Eq.~(\ref{eq:U1_metal_new}) and discussed in the previous section
can be applied to deal with quasi-degenerate eigenvalues here. Again, the matrix~$\vec{U}^{(1,\delta)}(\vec{k})$ appearing in Eq.~(\ref{eq:polar_bis}) is associated with a perturbation along the Cartesian axis~$\delta$, while the $\vec{\Omega}^{(\gamma)}(\vec{k})$ matrix is associated 
to a perturbation along the axis~$\gamma$.

\section{Validation and Performance}
\label{sec:results}
To validate our implementation we show how our simulations converge with respect to the numerical
parameters used in the calculation in Sec.~\ref{sec:polar_convergence}.
Furthermore, we compare our PT polarizabilities to those obtained from finite differences 
in Sec.~\ref{sec:polar_with_fd}. These tests are then extended
to periodic systems in Sec.~\ref{sec:dielectric}. The computational performance of the implementation is discussed in Sec.~\ref{sec:scaling}.

\subsection{Convergence with respect to Basis Set Size and $\vec{k}$-Point Grid Density \label{sec:polar_convergence}}
We observe that our calculated polarizability tensors are most sensitive to the basis set size and the amount of $\mathbf{k}$-points used in the simulation, as shown below. All other numerical parameters either influence the results very little or show a similar convergence behavior as in ground-state DFT calculations. 

First, we discuss the convergence of polarizabilities in our implementation with respect to the basis set size used for the expansion of the Kohn-Sham states in Eq.~(\ref{eq:expansion}). As an example, we use ethylene~(C$_2$H$_4$), for which we compute the three diagonal components~$\alpha_{\gamma\gamma}$ of the polarizability tensor using LDA~\cite{Perdew/Zunger:1981,Ceperley/Alder:1980}. In all cases, the PT calculations were performed for the same geometry, i.e., the structure obtained by geometry optimization~(maximum residual force~$<10^{-4}$~eV/$\mbox{\AA}$) with $tight$ basis sets and numerical settings. The C-C bond of the molecule is oriented along the Y-axis.

\begin{figure}
 \centering
 \hfill
 \includegraphics[width=0.45\textwidth]{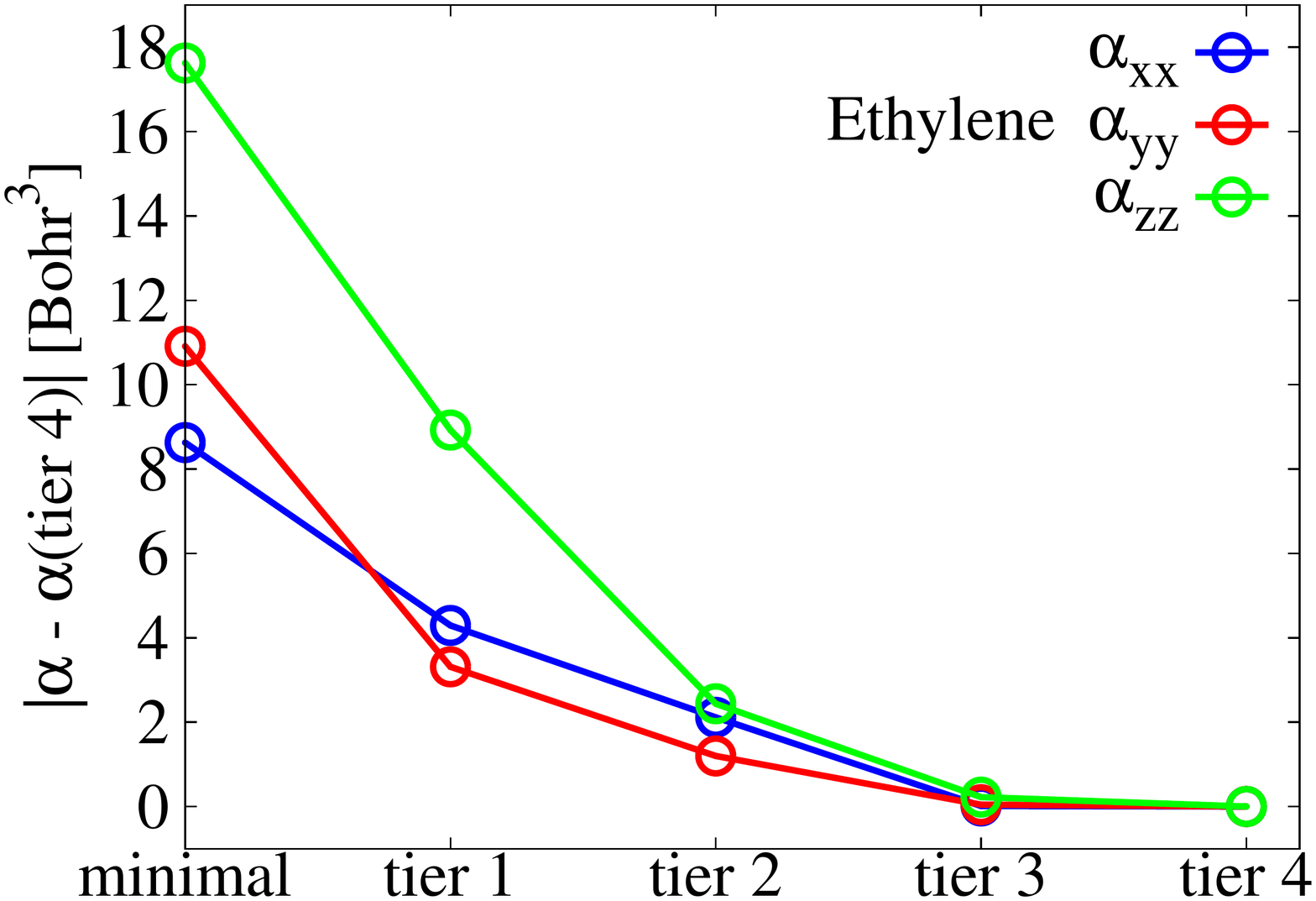}
 \hfill
 \includegraphics[width=0.45\textwidth]{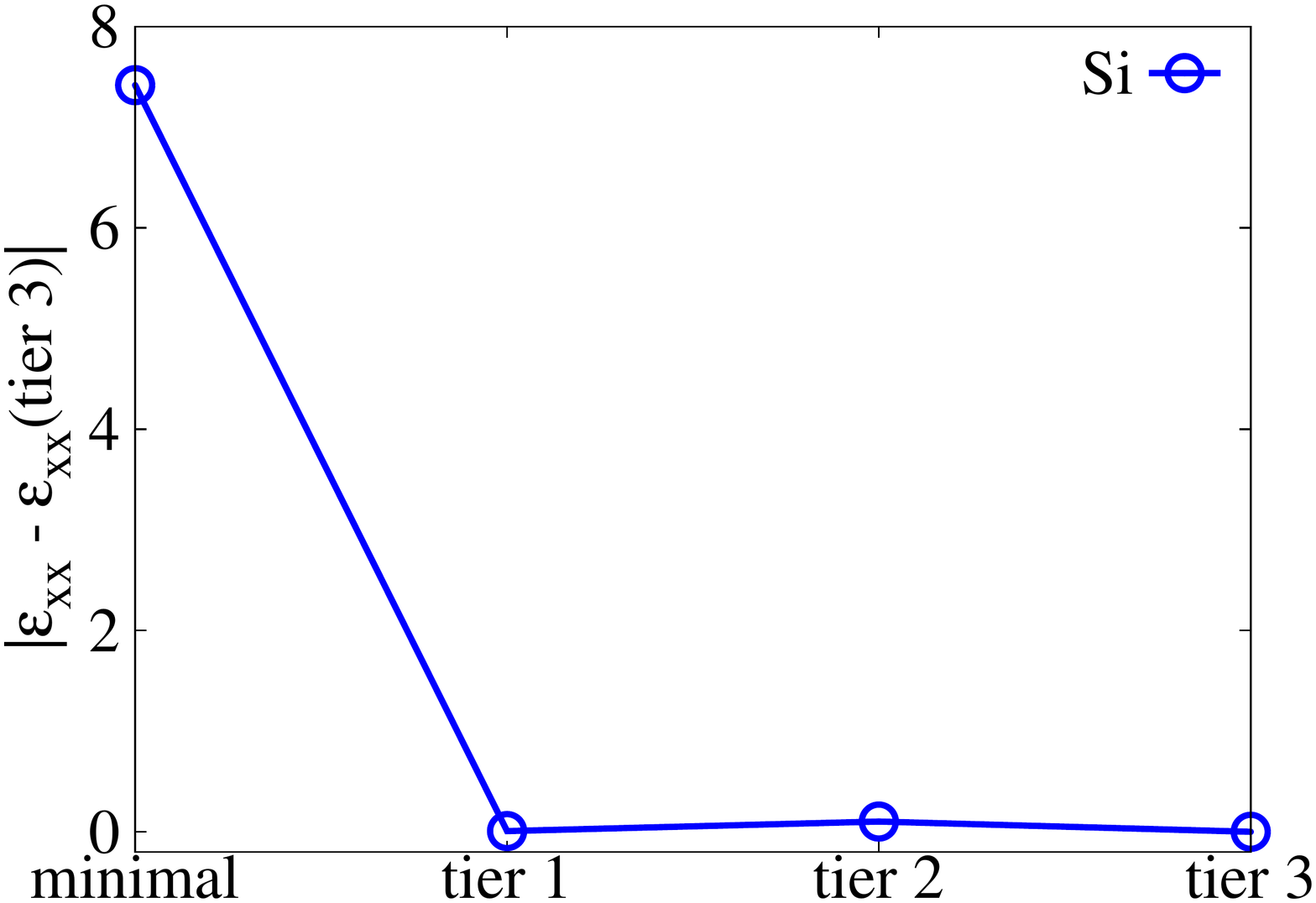}
 \hfill
 \caption{
Convergence behaviour of the polarizabilities~$\alpha_{xx}, \alpha_{yy}, \alpha_{zz}$ of ethylene and of the high-frequency dielectric constant~$\varepsilon_{xx}^\infty$ of bulk silicon~($16 \times 16 \times 16$ $\vec{k}$-points) with respect to the basis set size~(see  text).
}
  \label{fig:c2h4_convergence_basis}
\end{figure}

Figure~\ref{fig:c2h4_convergence_basis} shows the absolute error in the diagonal components of the polarizability tensor with increasing basis-set size.  Here, a minimal basis includes exactly one basis function per electron; additional functions are then added in groups, so-called \textit{tier 1}, \textit{tier 2}, etc., basis sets (see Ref.~\cite{Blum2009} for more details). The polarizabilities converge slowly with the basis set size in finite molecular systems as ethylene: Although getting qualitatively correct results, the maximum absolute~(relative) error is for instance still  $2.44$~$a_0^{3}$~($11$~\%) at a \textit{tier 2} level. Only at the \textit{tier 3} level we get a maximum absolute~(relative) error of $0.23$~$a_0^{3}$~($1$~\%).
For semi-infinite systems, the dielectric constant, which is directly proportional to the polarizability as noted in Eq.~\ref{eq:dielec}, converges much faster with increasing basis set size, as also shown in Fig.~\ref{fig:c2h4_convergence_basis} for bulk silicon. Even at a \textit{tier 1} level we essentially achieve convergence with an absolute~(relative) error of $0.007$~($0.05$~\%).

\begin{figure}[h]
 \centering
 \includegraphics[width=0.4\columnwidth,angle=5]{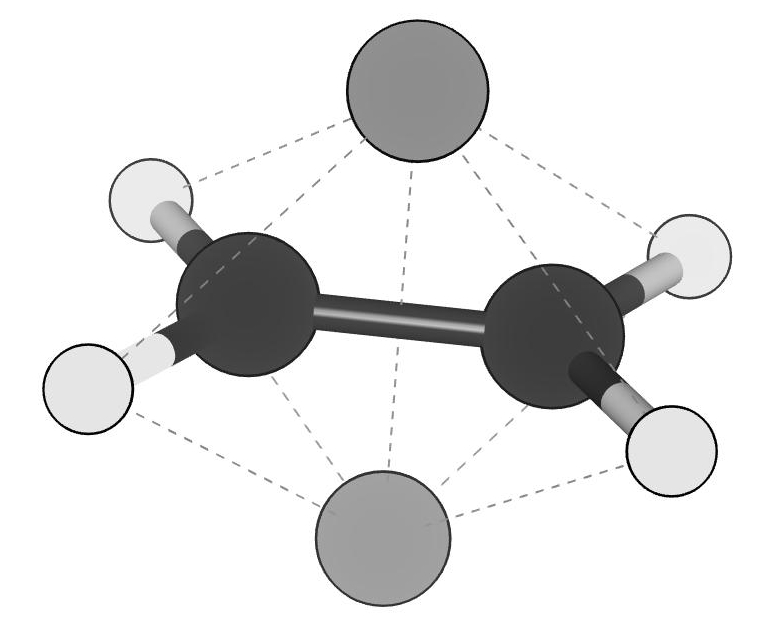}
 \caption{Sketch of the C$_2$H$_4$ molecule and its two ghost atoms used to improve the convergence. Ghost atoms are pictured at the top and bottom of the molecule from this perspective.}
  \label{fig:c2h4_ghost}
\end{figure}

The slower convergence observed for molecular systems arises from the inhomogeneous distribution of the localized basis sets in isolated systems. The standard basis sets in FHI-aims have been optimized to obtain converged ground state energies, but are not necessarily even-tempered for the calculation of polarizabilities, which can create an imbalance in the extent of the polarization that is possible in different directions.  One possibility to improve convergence would be the construction of basis sets that are specifically tailored for the calculation of polarizabilities, see for instance Ref.~\cite{Rappoport_Furche_JCP2010} for an example of basis sets adapted for polarizabilities, Ref.~\cite{hyperpolarizability_basis_convergence} for hyperpolarizabilities, or Ref.~\cite{Laasner:fx} for magnetic response properties. Alternatively, it is possible to include extra basis functions in otherwise empty regions to span the space much more efficiently. As shown below, this allows to reduce the computational cost by using much smaller overall basis sets without sacrificing accuracy. The difficult task in this procedure is to determine in which region of space the original basis sets are not sufficient, in order to determine where to best place the extra basis functions. In general, the symmetries of the molecule are helpful in this task and thus need to be considered as well. We illustrate this procedure for the polarizabilities of the C$_2$H$_4$ molecule with LDA (see Table~\ref{tab:ghost_lda}). It is clear that the addition of 2 carbon-like ghost atoms (i.e. only the \textit{tier 1} basis set of a carbon atom), which we positioned below and above the molecular plane on the bisection of the C-C segment (see Fig~\ref{fig:c2h4_ghost}), significantly improves the convergence, almost to the level of \textit{tier 2}, but at only half the computing time. Please note that simply increasing the onset of the cutoff potential for the usual basis sets in FHI-aims does not improve the performance of our results.

\begin{figure}[h]
 \centering
 \includegraphics[width=0.9\columnwidth]{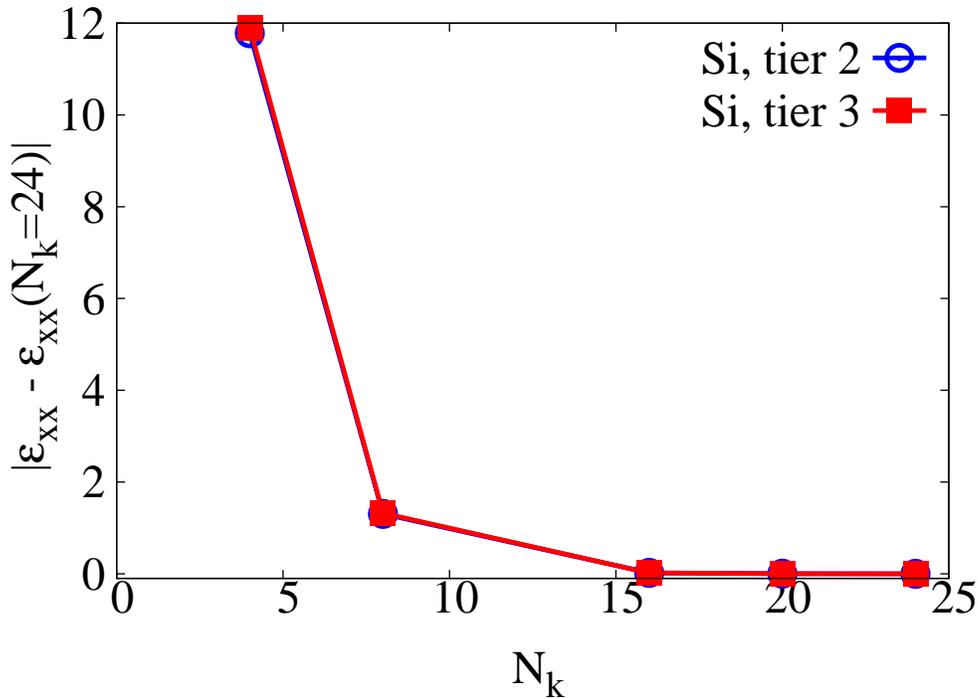}
 \caption{Convergence of the diagonal components of the high-frequency dielectric constant~$\varepsilon_{xx}^\infty$ of bulk silicon with respect to the $\mathbf{k}$-point density. The size of the $\mathbf{k}$-grid is $N_{k}\times N_{k} \times N_{k}$. \textit{Tight} grid settings and \textit{tier 2} as well as \textit{tier 3} basis sets are used. The benchmark value is calculated using $N_{k}$=24.}
  \label{fig:Si_convergence_k_point}
\end{figure}

\begin{table}[h]
\center
 \begin{tabular}{l|rrrr|r}\hline\hline
 basis set & \multicolumn{4}{c|}{polarizability} & time (s) \\ \hline
 tier1 & 21.8 & 32.8 & 13.2 & (22.6) & 7.4   \\ 
 tier1 + 2 H-ghosts & 22.3 & 33.2 & 18.6 & (24.7) & 12.9  \\ 
 tier1 + 2 C-ghosts & 24.4 & 33.5 & 19.6 &(25.8) & 18.4  \\ 
 tier2 & 23.9 & 35.0 & 19.7 & (26.2) & 36.3 \\ 
 \hline\hline
\end{tabular}
\caption{Influence of H- and C-like ``ghost'' atoms on the diagonal elements of the polarizability of C$_2$H$_4$, using light settings and  LDA.  Numbers in brackets indicate the mean polarizability.}
\label{tab:ghost_lda}
\end{table}
Finally, to study the sensitivity of the polarizability tensor on the $\mathbf{k}$-point grid density in periodic systems, we also use silicon as example. Fig.~\ref{fig:Si_convergence_k_point} displays the convergence behavior  with respect to the size of the reciprocal-space $\vec{k}$-mesh in the primitive Brillouin zone. We observe a maximum absolute~(relative) error of $0.12$~($0.15$~\%) when using $16 \times 16 \times 16$ $\vec{k}$-points with respect to the converged result. This convergence behavior is comparable or slightly slower than what is observed for the total energy.

\subsection{Validation against Finite Differences}
\label{sec:polar_with_fd}
To validate our PT implementation, we also compared the obtained polarizabilities of 32 selected molecules to the ones obtained via finite difference calculations, as detailed in the Appendix. There, the details for each individual molecule can be found; here, this data is succinctly summarized in Tab.~\ref{tab:mae},  where we list the mean absolute percentage error~(MAPE) and the mean absolute error~(MAE) for all tested molecules. Overall, we find an excellent agreement between our implementation and the finite-difference results. 

\begin{table}[h]
\centering
\begin{tabular}{c |  c c }
\hline \hline
$\vert\alpha^\t{FD}-\alpha^\t{PT}\vert$         &   MAE~($a_0^{3}$) &  MAPE \\
\hline 
Dimers &    0.0004  & 0.0007\%   \\
Trimers &    0.0002        & 0.001 \% \\
Molecules & 0.0002 & 0.0008 \% \\
\hline \hline
\end{tabular}
\caption{
Mean absolute error (MAE) and mean absolute percentage error (MAPE) for the difference between the polarizabilities obtained via PT and finite differences~(FD) for a set of 16 dimers, 5 trimers, and 11 molecules. All calculations are performed at the LDA level of theory with fully converged
numerical settings and relaxed geometries. Detailed informations including the values for each individual molecule can be found in the Appendix.}
\label{tab:mae}
\end{table}

\subsection{Extended Systems: High-Frequency Dielectric Constant}
\label{sec:dielectric}
In order to validate our implementation for extended systems, we have calculated the dielectric constant of several semiconductors using the local-density approximation~(LDA~\cite{perdew-1992}) and  
the generalized gradient approximation~(GGA-PBE~\cite{Perdew-1,Perdew-2}) and compared it with experimental and theoretical data compiled from literature~\cite{Giannozzi1991,DalCorso1996}, see Tab.~\ref{tab:compare_with_ref}. 
All calculations have been performed at the theoretical equilibrium lattice constant using 16$\times$16$\times$16~$ \mathbf{k}$-points in the primitive unit cell and ``tight'' basis set and integration settings.
Also, we list LDA/GGA literature results obtained using a plane wave basis set and norm-conserving pseudopotentials~(NCPP)~\cite{Giannozzi1991,DalCorso1996} or the projector augmented wave method ~(PAW~\cite{Gajdo2006,Petousis2016}).
With respect to experiment, we note that all LDA and GGA calculations overestimate the electronic dielectric constant by roughly~$10$\% due to the well-known fact that these functionals yield too small band gaps~\cite{Gironcoli1989,Giannozzi1991}. 
 
With respect to theoretical results, the most recent literature data computed with the PAW method (LDA~\cite{Gajdo2006}; PBE~\cite{Petousis2016}) is in excellent agreement with our implementation.
Slightly larger deviations are observed with respect to earlier calculations that rely on norm-conserving pseudopotentials~(NCPP): The agreement is generally better with literature results obtained
using non-linear core corrections~\cite{Louie1982,DalCorso1993}. For instance, this can be observed for GaSb: The work of Dal Corso {\it et al.}~\cite{DalCorso1996} made use of non-linear core corrections, 
but not the earlier one of Giannozzi {\it et al.}~\cite{Giannozzi1991}. For the latter work, the use of a smaller $\vec{k}$-point grid may also be partially responsible for the observed deviations.

\begin{table}[t]
\centering
\begin{tabular}{l||r||r|r||rrr|r}
\hline \hline 
        & \multirow{3}{*}{\parbox{0.075\linewidth}{\centering Exp.\\\cite{Philipp1963-Si-GaP, Beer1968-AlP-1,Monemar1970-AlP-2,Fern1971-AlAs,Hass1962-AlSb-GaSb}}}      &\multicolumn{2}{c||}{this work}      & NCPP & NCPP & PAW  & PAW \\
        &       &\multicolumn{2}{c||}{(all}           & 1991 & 1996 & 2006 & 2016 \\
        &       &\multicolumn{2}{c||}{electron)}      & \cite{Giannozzi1991} & \cite{DalCorso1996} & \cite{Gajdo2006} & \cite{Petousis2016} \\\cline{3-8}
        &       & LDA  & PBE  & \multicolumn{3}{|c|}{LDA} & PBE\\\hline
Si	& 12.1	& 13.2 & 12.9 & 13.6 & -   & 13.3  &  13.1  \\
AlP 	&  7.5	&  8.4 &  8.2 &  -   & 8.2 &  8.3  &   8.1  \\
AlAs	&  8.2	&  9.5 &  9.5 &  9.2 & 9.3 &   -   &   9.5\\
AlSb	& 10.24	& 11.7 & 11.9 & 12.2 &11.4 &   -   &  12.1\\
GaP 	&  9.0	& 10.6 & 10.6 &  -   &10.0 &   -   &  10.6\\
GaSb	& 14.44	& 16.0 & 15.5 & 18.1 &16.7 &   -   &    - \\\hline\hline
\end{tabular}
\caption{Comparison of the high-frequency dielectric constants of various semiconductors computed at the LDA/PBE level with literature values: Tight-default settings and basis sets as well as a 16$\times$ 16$\times$ 16 $\mathbf{k}$-point mesh are used.}
\label{tab:compare_with_ref}
\end{table}

\subsection{Performance and Scaling of the Implementation}
\label{sec:scaling}

\begin{figure}[ht]
\centering
 \hfill
 \includegraphics[width=0.45\textwidth]{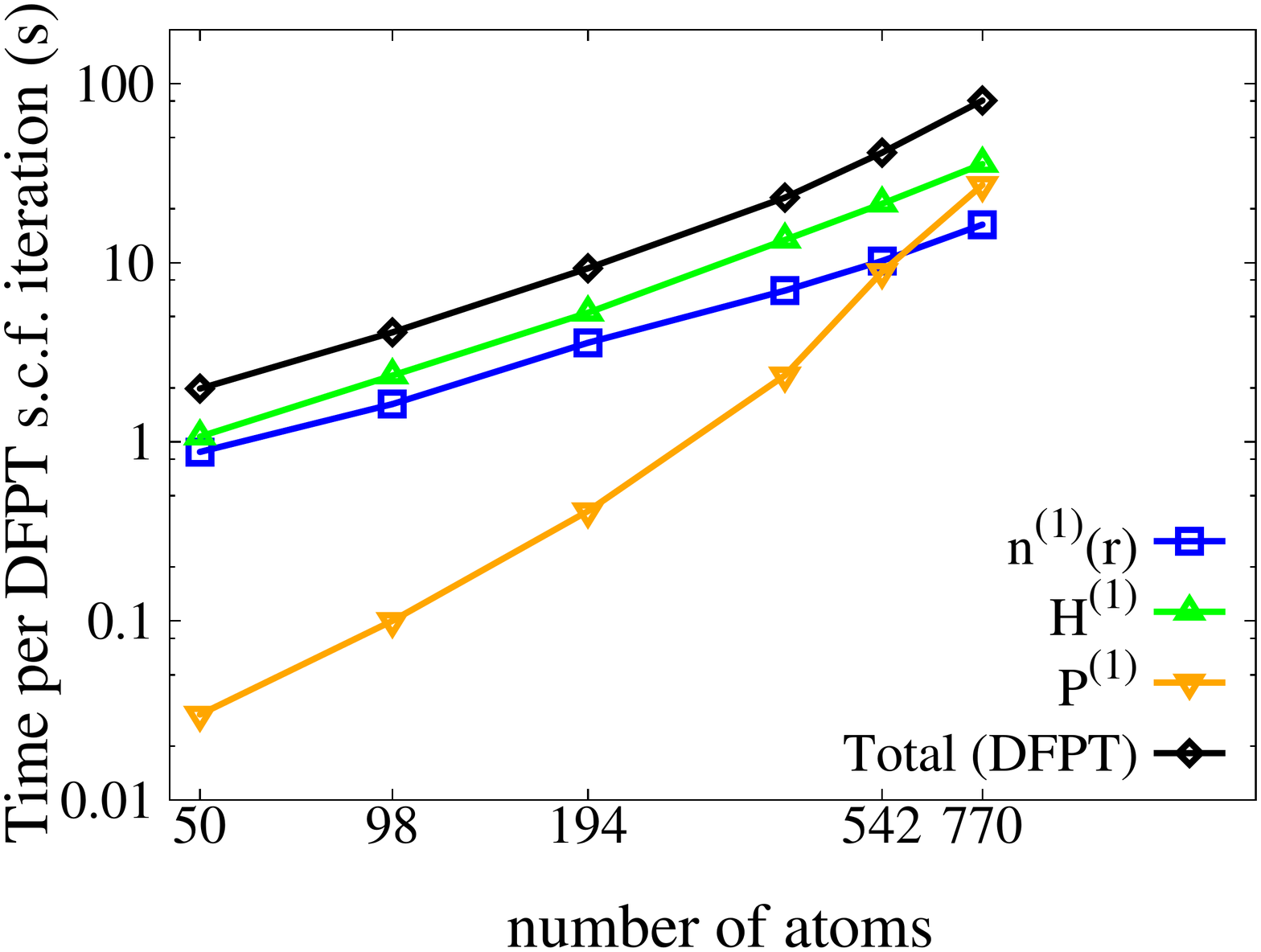}
 \hfill
 \includegraphics[width=0.45\textwidth]{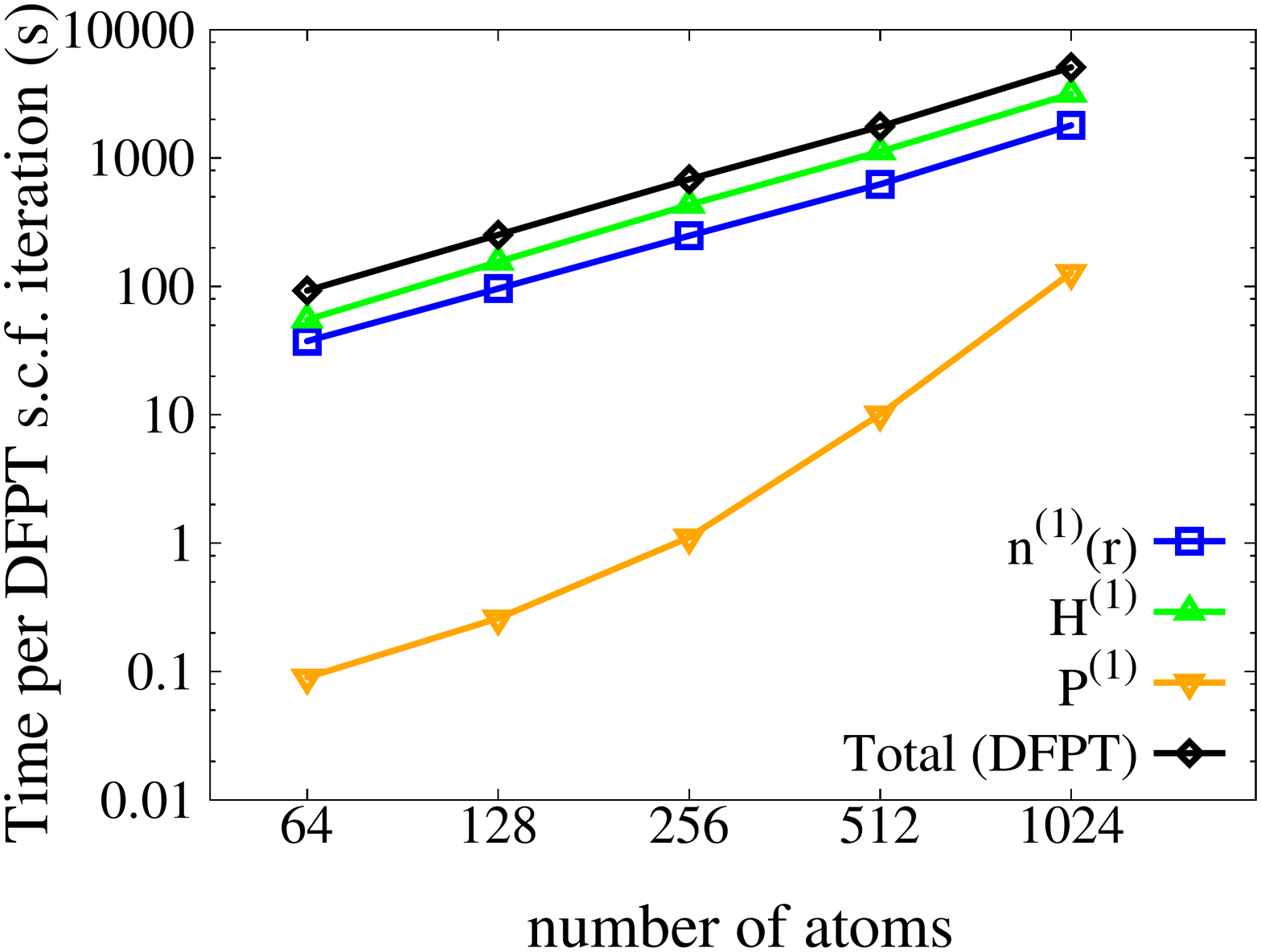}
 \hfill
 \caption{H(C$_2$H$_4$)$_n$H molecules: 
CPU time per PT cycle required for finite H(C$_2$H$_4$)$_n$H molecules~(left) and periodic diamond~(right) as a function of the number of atoms~(diamond: in the unit cell) on 32 cores~(see text). Following the flowchart in Fig.~\ref{fig:DFPT_flowchart_E_filed}, also the timings required for the  computation of the individual responses,~i.e.,~the ones of the density $n^{(1)}(\vec{r})$, of the Hamiltonian matrix $\mathbf{H}^{(1)}(\vec{k})$, and of the density matrix $\mathbf{P}^{(1)}$, are given.}
 \label{fig:scaling_c2h4_mol}
\end{figure}

To demonstrate the performance and scaling of our implementation, we show timings for the H(C$_2$H$_4$)$_n$H molecules with variable $n=8-128$ and diamond. In the latter case, different supercell sizes were considered by increasing the number of building units in the unit cell from~(C$_2$)$_{32}$ to (C$_2$)$_{512}$. All calculations use light settings and the LDA functional. Only the $\Gamma$ point is considered in the periodic case.
Calculations were performed on a single node featuring two Intel Xeon E5-2698v3 CPUs~(32 cores) and $4$~Gb of RAM per core.

\begin{table}
\centering
\begin{tabular}{c |  c | c   }
\hline \hline    
                   & H(C$_2$H$_4$)$_n$H & Diamond \\
\hline
$n^{(1)}(\vec{r})$          & 1.1   &          1.4 \\
$\vec{H}^{(1)}(\vec{k})$          & 1.4   &   1.5       \\ 
$\vec{P}^{(1)}$          & 2.5   &   2.6    \\
\hline
Total              & 1.3   &  1.4      \\
\hline \hline 
\end{tabular}
\caption{Fitted CPU time exponents~$\alpha$ for the H(C$_2$H$_4$)$_n$H molecules (n=8-128) and 
the periodic diamond  discussed in the text. The fits were performed using the expression~$t = c N^{\alpha}$ for the CPU time as function of the number of atoms~$N$.}
\label{tab:scaling}
\end{table}

For the timings shown in Fig.~\ref{fig:scaling_c2h4_mol} (molecules), we find that
the integration of the Hamiltonian response matrix~$\vec{H}^{(1)}(\vec{k})$ determines the computational time for 
small system sizes,~i.e.,~for less than 200 atoms. Like for the update of the response
density~$n^{(1)}$, which involves similar numerical operations, we find a scaling of nearly~$O(N)$ for
this step~(see Tab.~\ref{tab:scaling}), as it is the case in ground-state DFT calculations~\cite{Blum2009}. 
For very large system sizes~($N\gg 1000$), the update of the response density matrix~$\vec{P}^{(1)}$ becomes dominant,
since it scales with~$O(N^{2.5})$ in this regime. As discussed in Sec.~\ref{sec:Implementation}, 
the computation of~$\vec{P}^{(1)}$ requires matrix multiplication operations, which traditionally scale~$O(N^3)$.
For bulk diamond we find a similar behavior and fit similar exponents, as shown in Fig.~\ref{fig:scaling_c2h4_mol} and Tab.~\ref{tab:scaling} as well. 
We note that the prefactors to these timings are higher for dense 3D systems than for 1D systems and that they also are system dependent.
Our real-space PT implementation thus exhibits a similar scaling as the underlying DFT calculations, as it is generally 
the case for DFPT/CPSCF codes.

\begin{figure}[ht]
\centering
 \includegraphics[width=0.9\columnwidth]{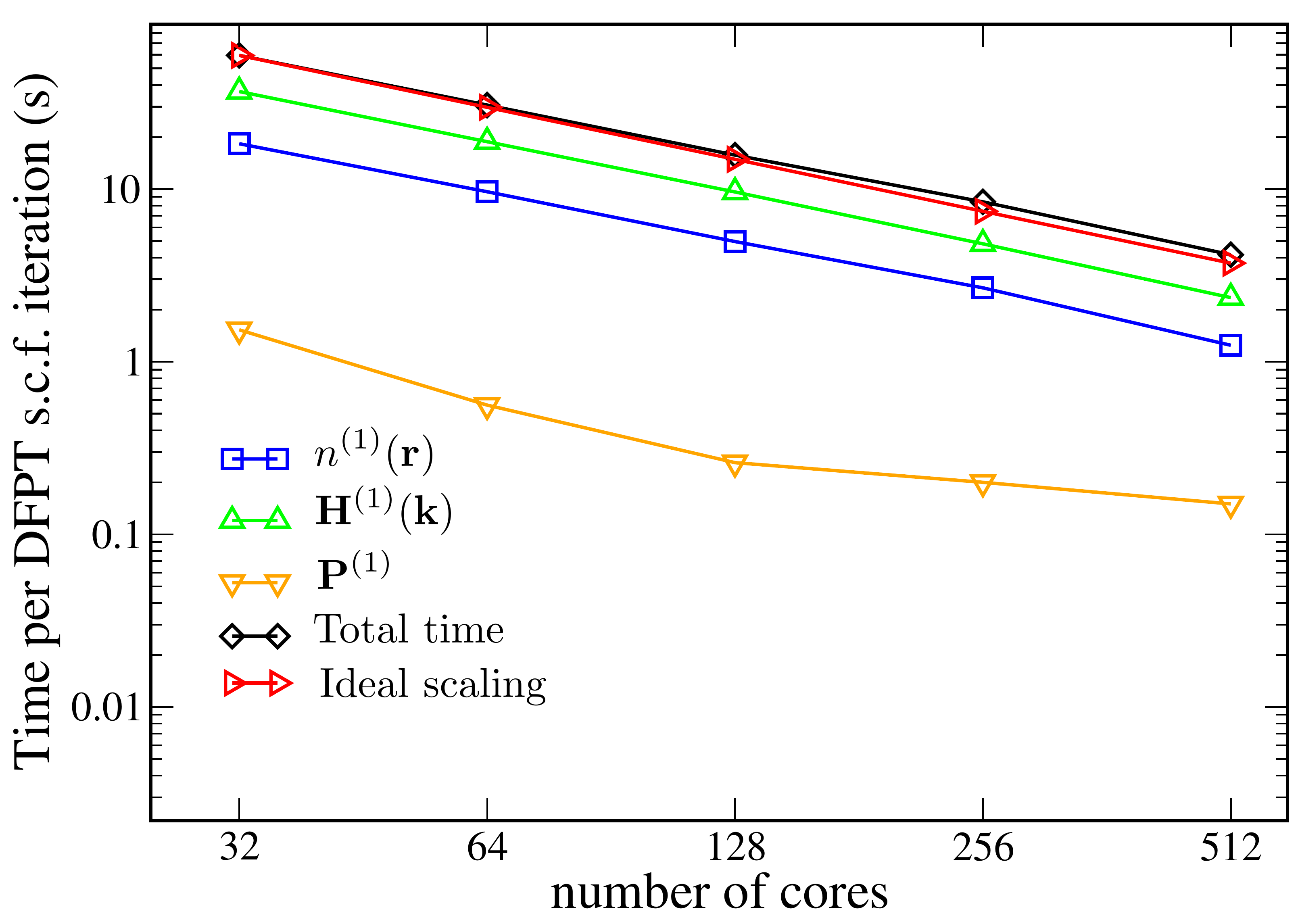}
 \caption{Scaling of the CPU time per PT cycle with the number of cores~(parallel scalability) for the paracetamol crystal (form II) containing 160 atoms in the unit cell. 
 \textit{Tier 1} basis sets and a $2\times 2\times 2$ $\vec{k}$-grid are used. The time required for the computation of the individual response properties is also shown.}
 \label{fig:paracet_scaling}
\end{figure}

In summary, we find an overall scaling behavior that is always smaller than $O(N^2)$ for the investigated system
sizes both in the molecular and the periodic case. 
Note that a parallelization over cores is already part of the presented implementation, given that the discussed real-space formalism closely follows the strategies used for the parallelization of ground-state DFT calculations in FHI-aims~\cite{Blum2009,Havu/etal:2009}. 
As shown in Fig.~\ref{fig:paracet_scaling} for a unit cell of the paracetamol crystal containing 160 atoms, almost ideal scaling is achieved for the time per PT iteration when different number of CPU-cores~(same specifications as in the previous tests) are used.
Still, it is very gratifying to see that even quite extended systems with more than 100 atoms in the unit cell are in principle treatable within the relatively moderate CPU and memory resources offered by a single state-of-the-art workstation.

\section{Application: Harmonic and Anharmonic Raman Spectra\label{sec:application}}

In order to showcase the usefulness and efficiency of our implementation, we calculate the non-resonant Raman spectra of paracetamol in its molecular form, as well as in its first (monoclinic) and second (orthorhombic) crystalline polymorph. More specifically, we wish to investigate the impact of anharmonicities on these spectra, due to their acknowledged importance in H-bonded, flexible systems~\cite{Mariana_PRL_2016, Putrino_PRL_2002, Warshel1971}. Focusing on molecular crystals, which often exist in multiple competing polymorphs with very different physicochemical properties, makes it necessary to have an accurate and efficient model to characterize such structures. Taking into account anharmonicities in the computation of Raman spectra is of crucial importance, as has already been proven in the past, for example for the characterization of phase transitions in high-pressure ice~\cite{Putrino_PRL_2002}.

Vibrational Raman spectra are typically computed in the harmonic approximation, where the Raman intensities are  proportional to the derivatives of the polarizability with respect to atomic displacements, as detailed, for example, in Refs.~\cite{Neugebauer_JCC2002,Veithen:2005gf}. In this work, we calculate these harmonic Raman intensities through finite-differences by numerically computing the derivatives of the polarizability tensor via finite displacements of the nuclear coordinates. These displacements are performed in the unit cell, since only phonons at the $\Gamma$ point of the lattice contribute to the Raman intensity. Additionally, we also compute \textit{anharmonic} Raman spectra through the calculation of polarizability autocorrelation functions in thermodynamic equilibrium. We simulate the nuclear dynamics using \textit{ab initio} molecular dynamics and compute the polarizabilities along these trajectories via PT. As explained in, e.g., Ref~\cite{Berne_Pecora}, the polarizability tensor $\bm{\alpha}$ can be divided into an isotropic $\bar{\alpha}$, and an anisotropic component $\bm{\beta}$,
\begin{eqnarray}
\bm{\alpha}&=&\bar{\alpha} \bm{I}+\bm{\beta} \\
\nonumber \bar{\alpha}&=&\frac13(\alpha_{xx}+\alpha_{yy}+\alpha_{zz}) ~,~
 \textrm{Tr}[\bm{\beta}]=0 ~.
\end{eqnarray}

As shown in Refs.~\cite{Putrino_PRL_2002, Pagliai_JCP2008} where pioneering simulations of this type were presented, the (non-resonant) Raman intensity $I(\omega)$ is then expressed as a sum of isotropic and anisotropic parts as 
\begin{eqnarray}
 I(\omega) &= & I_\t{iso} +\frac43 I_\t{aniso} \label{eq:raman-int}\\
\nonumber I_\t{iso} & = &\frac{N}{2\pi} \int_{-\infty}^{+\infty}dt \e^{-i\omega t} \langle\bar{\alpha}(0)\bar{\alpha}(t)\rangle  \\
\nonumber I_\t{aniso} & = &\frac{N}{2\pi} \int_{-\infty}^{+\infty}dt \e^{-i\omega t} \frac{1}{10}\langle \textrm{Tr}[\bm{\beta}(0)\cdot \bm{\beta}(t)]\rangle ~.
 \end{eqnarray}
Here, $N$ is again the number of atoms in the system.
Furthermore, since the autocorrelation functions~$\langle\cdot\rangle$ are computed classically, a quantum correction factor is usually applied. Due to the fact that the classical correlation function better approximates the Kubo transform of the quantum autocorrelation function, we multiply $I(\omega)$ in Eq. \ref{eq:raman-int} by $\beta \hbar \omega/(1-\e^{-\beta \hbar \omega})$, where $\beta=1/k_\t{B}T$ ~\cite{Ramirez_2004}. Further frequency-dependent factors that multiply the vibrational Raman lineshapes are experiment-dependent \cite{McCreery_2005,Keresztury_1993,Chalmers-Griffiths_2001}. Here, we normalized experimental and theoretical spectra by their areas for comparison.
All MD trajectories used in this paper have been obtained using the PBE functional in combination with many-body van der Waals interactions~\cite{Distasio_Tkatchenko_2014} (PBE+MBD), which have been previously shown to play an important role for the accurate assessment of  potential-energy surfaces~(PES) and free energies of molecular crystals~\cite{Pagliai_JCP2008,Mariana_PRL_2016,ReillyTkatchen2014,HojaTkatchen2017}.

\begin{figure}[ht]
 \includegraphics[width=1.0\columnwidth]{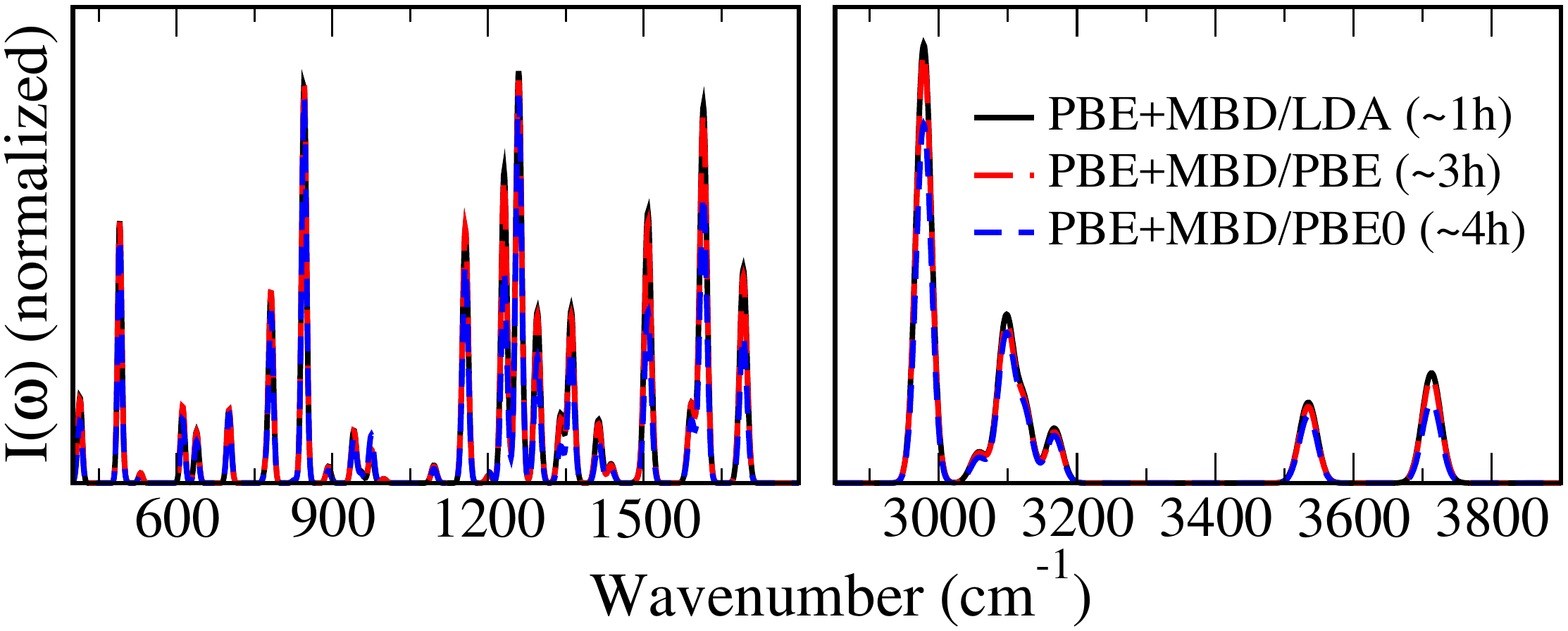}
 \caption{Harmonic Raman spectrum of the paracetamol molecule: The notation XX/YY denotes that the PES~(energy and forces) were calculated with the XX functional, while the YY functional was used for the polarizabilities in the PT part. In this figure, the PES is always
obtained at the PBE+MBD level, while different functionals are used for the polarizabilities. \textit{Tight} settings and basis sets were used and the calculated (finite-difference) Raman intensities were convoluted with a Gaussian function of fixed width for better visualization.  Computational time required for each simulation is also denoted.}
 \label{fig:ha_func_polar}
\end{figure}

We first analyze the sensitivity of the harmonic Raman intensities of the paracetamol molecule to the employed exchange-correlation functional. In Fig.~\ref{fig:ha_func_polar}, the PES is always treated at the PBE+MBD level, but different functionals are used to calculate the polarizabilities. 
We observe that the Raman spectra are essentially insensitive to the choice of xc-functional~(LDA, PBE, and hybrid functional PBE0) used in the PT part. This is due to the fact that the magnitude of the Raman peaks are proportional to the polarizability derivatives with respect to atomic displacements and these derivatives are very similar in all functionals. Since evaluating the polarizabilities at the PBE0 level is four times more expensive than with LDA, we can decrease the cost of these simulations without sacrificing accuracy by evaluating the PT portion at the LDA level.

\begin{figure}[ht]
  \includegraphics[width=\columnwidth]{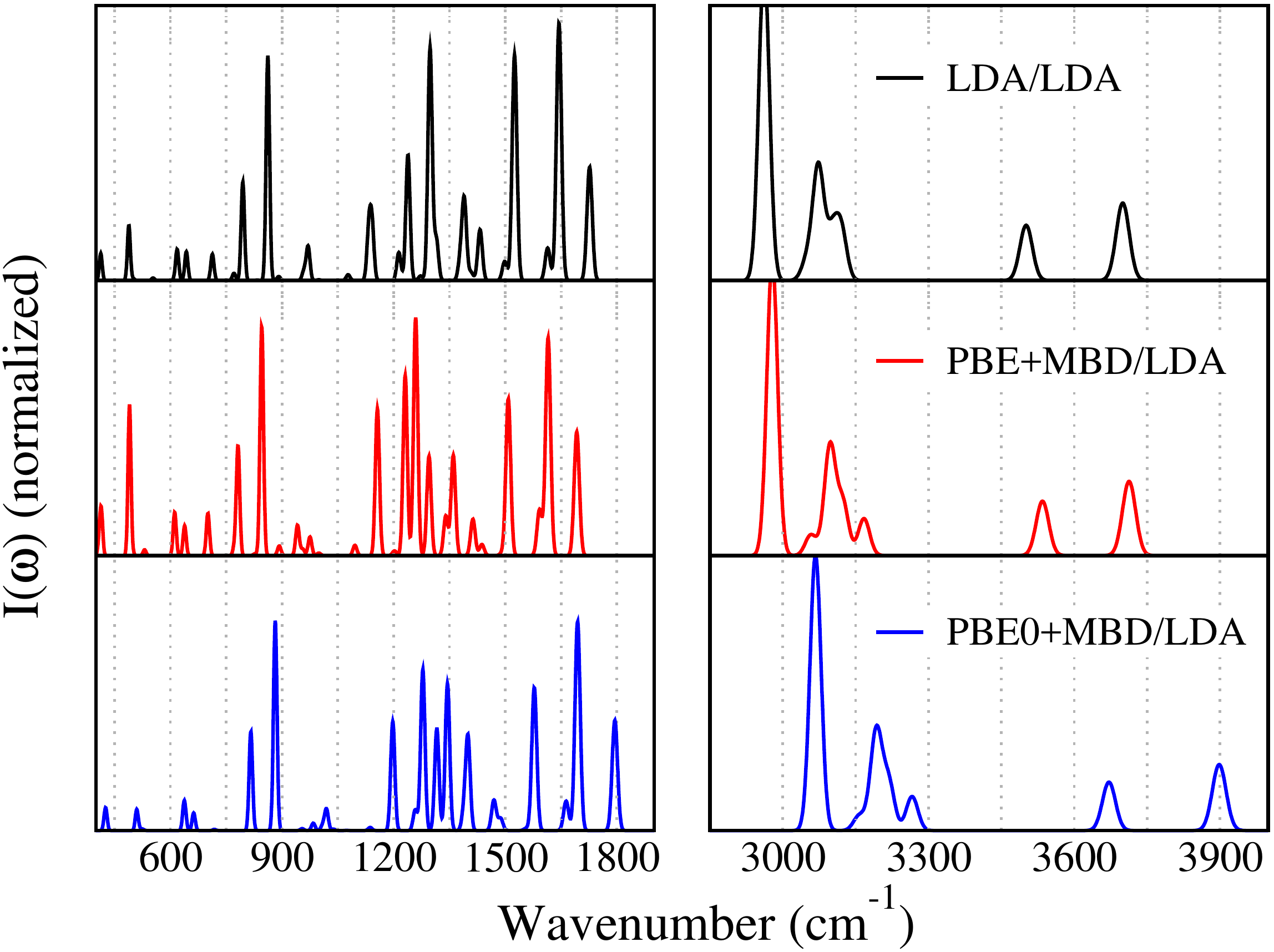}
  \caption{Harmonic Raman spectrum of the paracetamol molecule: The notation XX/YY denotes that the PES~(energy and forces) were calculated with the XX functional, while the YY functional was used for the polarizabilities in the PT part. In this figure, the polarizabilities 
  are always obtained at the LDA level, while different functionals are used for the PES. \textit{Tight} settings and basis sets were used and the calculated (finite-difference) Raman intensities were convoluted with a Gaussian function of fixed width for better visualization.}
  \label{fig:ha_func_pes}
\end{figure}

Conversely, the xc-functional chosen for the assessment of the PES has a large impact on the position of the peaks. In Fig.~\ref{fig:ha_func_pes}, we highlight this fact by showing the harmonic Raman spectra of the paracetamol molecule obtained using the LDA xc-functional for the polarizabilities, but different xc-functionals for the PES~(energy and forces including full geometry relaxation). Switching from LDA to PBE (or to PBE0) for probing the PES results in noticeable changes in the harmonic Raman spectrum, as can be seen from the shifts in the peak positions. We also note that the main differences introduced by Hartree-Fock exchange in the spectrum are rigid blue shifts of the peak positions, especially above 1000 cm$^{-1}$, which means that these vibrational modes become more stiff~\cite{RossiBlum2014, BaldaufRossi2015}.

\begin{figure}[ht]
\subfigure[]{
 \includegraphics[width=1.0\columnwidth]{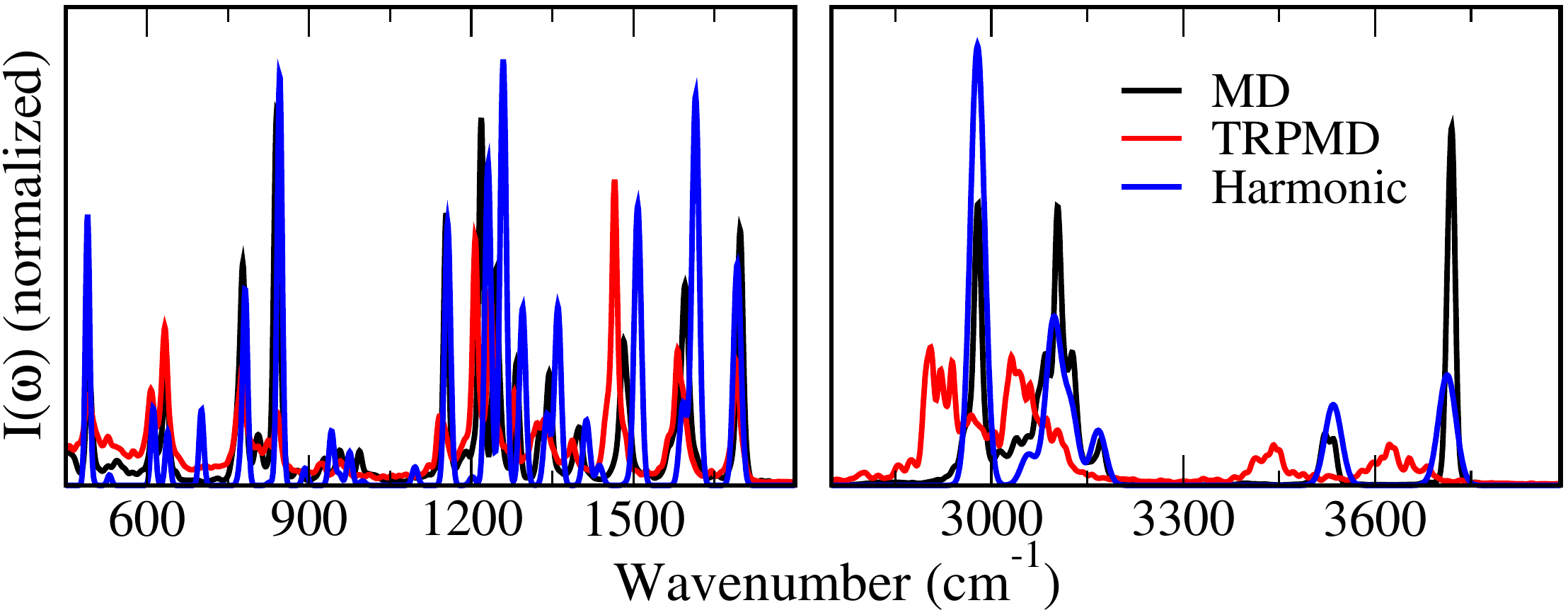}
}
\subfigure[]{
\includegraphics[width=1.0\columnwidth]{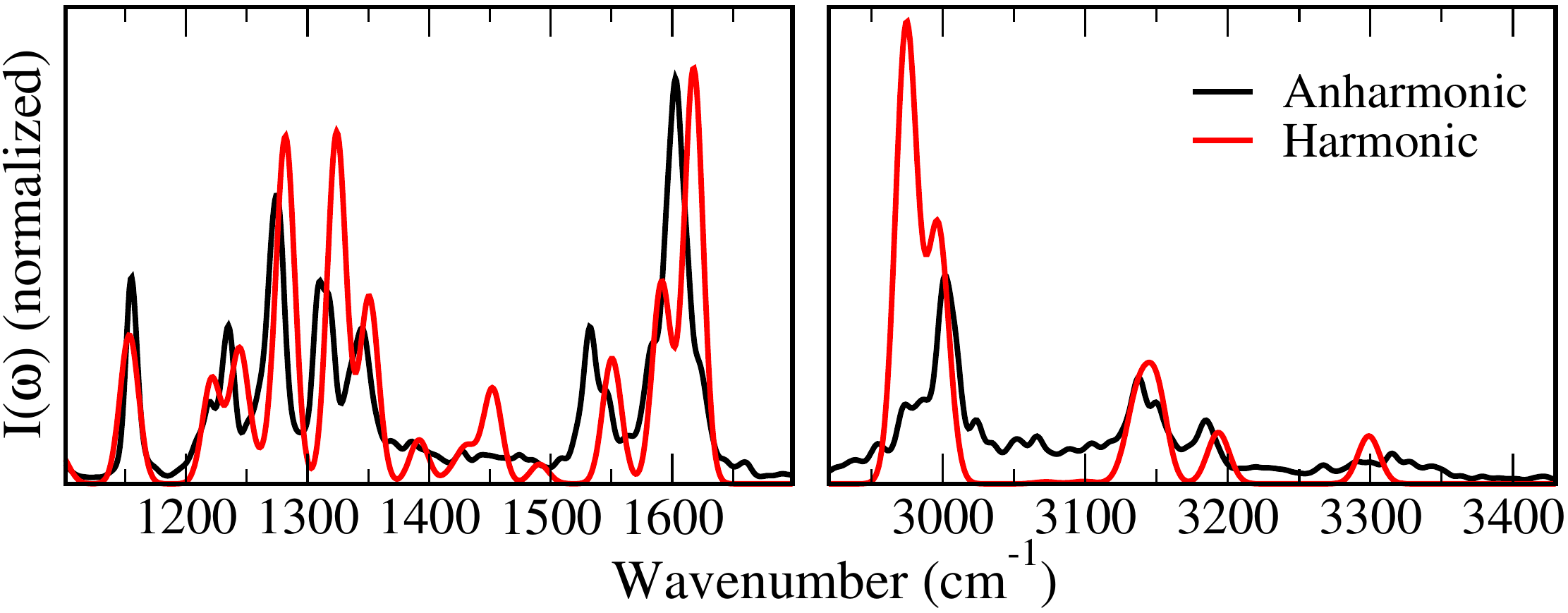}
}
 \caption{Comparison of harmonic and anharmonic (300 K) Raman spectra of (a) the paracetamol molecule, and (b) the paracetamol crystal (form I). In (a) we also show a spectrum obtained from thermostatted ring polymer molecular dynamics (TRPMD), which accounts for the quantum nature of the nuclei.  In all cases, the PES was probed with the PBE+MBD functional, while the polarizabilities were calculated with the LDA functional. Harmonic Raman intensities were convoluted with Gaussian functions for better visualization.}
 \label{fig:para_formI_ha_anha}
\end{figure}

In Fig.~\ref{fig:para_formI_ha_anha} (a) and (b), we show our calculated harmonic and anharmonic Raman spectra for the isolated paracetamol molecule and the paracetamol crystal in its monoclinic form I. For the molecule, we show anharmonic spectra obtained from an \textit{ab initio} MD trajectory at 300 K with classical nuclei and also spectra obtained from a thermostatted ring polymer MD trajectory at 300 K, which accounts for quantum-nuclear effects in the dynamics~\cite{RossiMano2014,RossiCeriotti2018}. We have used 16 replicas of the system and the Generalized-Langevin Equation thermostat proposed in Ref.~\cite{RossiCeriotti2018} for the internal modes of the ring polymer. While the general shape of the harmonic and anharmonic spectra are similar~(both for the molecule and the crystal), several peaks are shifted with respect to one another and feature different relative magnitudes, which substantiates the importance of anharmonic effects. In the molecular case, nuclear quantum effects induce a red-shift (with respect to the anharmonic classical spectrum) of about 70-100 cm$^{-1}$ in the high-frequency range. The effect is much less pronounced in the lower frequency regions, as expected~\cite{BaldaufRossi2015}. In the crystal, the lineshapes of the harmonic and anharmonic approximations are quite different, which highlights the fact that in our anharmonic spectrum we are able to capture the Raman peak lifetimes, while in the harmonic approximation we are simply convoluting the Raman intensities with Gaussian functions of a fixed width (and not explicitly calculating lifetimes). For periodic and condensed phase systems, we have previously shown \cite{RossiMichaelides2015,  RossiCeriottiJCPcomm, RossiCeriotti2018} that nuclear quantum effects would have a similar impact on the spectrum as for the molecular case. For water at room temperature for instance, the OH-stretch peaks are red-shifted by 150 cm$^{-1}$ solely due to nuclear quantum effects \cite{RossiCeriottiJCPcomm, RossiCeriotti2018}.

\begin{figure}[h]
 \includegraphics[width=1.0\columnwidth]{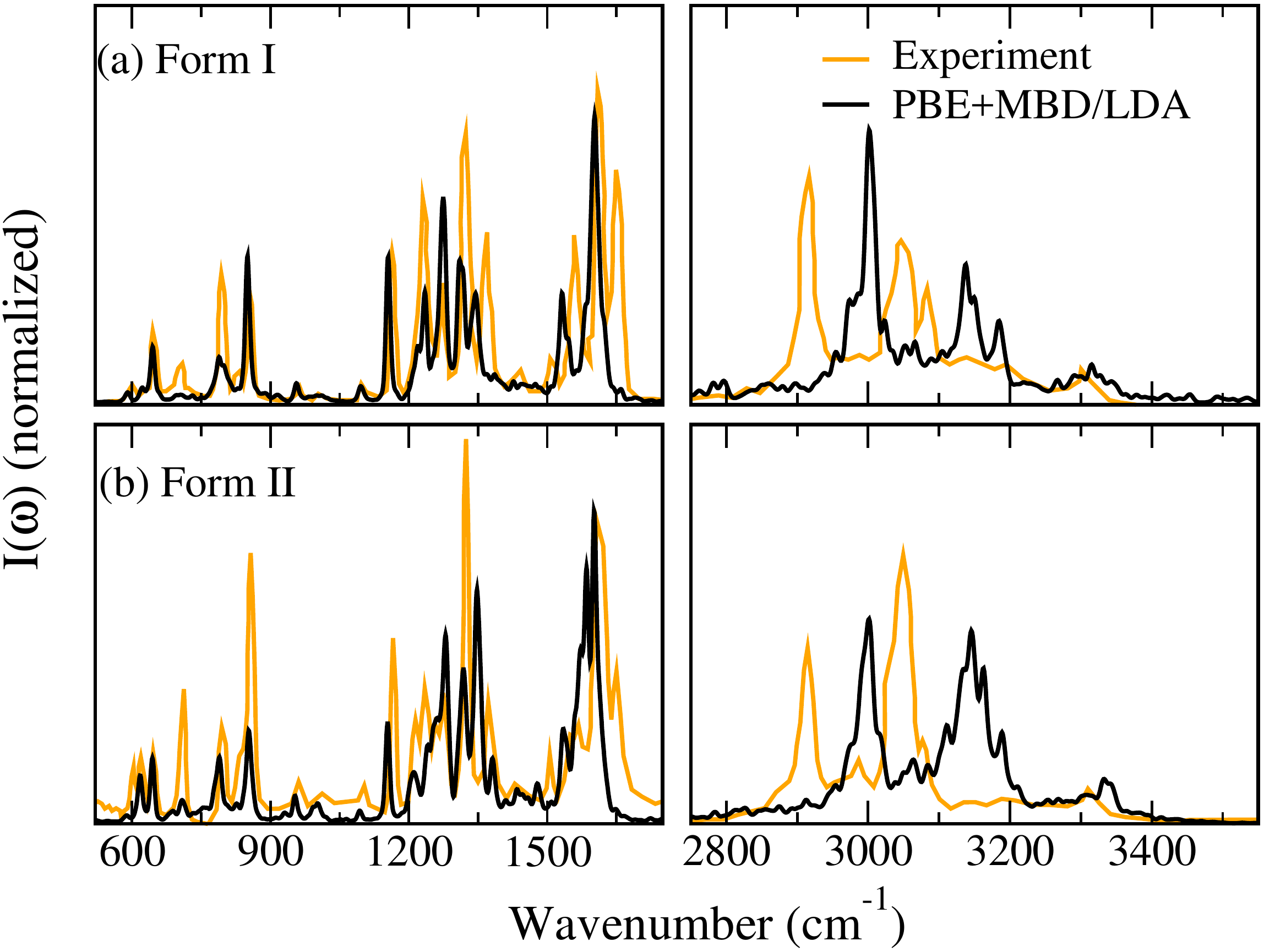}
 \caption{Raman spectra of paracetamol-form I (top) and II (bottom) calculated at 300 K.
Experiment from Ref.~\cite{Nanubolu} at room temperature. The spectra have been normalized to one in each panel.}
 \label{fig:formI-II_th-exp}
\end{figure}

In order to further evaluate the quality of our simulations we turn to a comparison to experimental data: Figure~\ref{fig:formI-II_th-exp} shows our computed anharmonic Raman spectra for polymorphs I and II of the paracetamol crystal, respectively, compared to experimental spectra from  literature. Both spectra were calculated from 2 independent MD runs of 15 picoseconds each. A time step of 0.5 femtosecond was used and the polarizability was computed every femtosecond.
Our results show a very good agreement with experiment, especially in terms of lineshapes for both crystalline forms. As previously discussed in literature and above, the observed rigid shifts between experimental and theroretical spectra originate from the choice of functional and the lack of nuclear quantum effects in the simulations. Employing a higher-level hybrid functional can be estimated to lead to blue-shifts of up to 180 cm$^{-1}$ for frequencies above 3000 cm$^{-1}$ (see Fig. \ref{fig:ha_func_pes}), while considering the quantum nature of the nuclei would red-shift these frequencies by up to 150~cm$^{-1}$ at high frequencies, as discussed above. To some extent, these effects hence cancel each other and are less pronounced at lower frequencies, which explains the good agreement observed in the 
600-1800 cm$^{-1}$ region between calculated and experimental spectra. However, the calculated spectra are still blue-shifted with respect to experiment above 2500 cm$^{-1}$, even though line-shapes are well reproduced. The inclusion of nuclear-quantum effects in the simulation would most likely solve this discrepancy, but the cost of such a simulation would be prohibitive at this point.

It is interesting to note that experimental spectra may sometimes differ slightly from one another as well. These differences are noticeable in the relative intensities of peaks or appearance/disappearance of low-intensity peaks~\cite{Nanubolu,Agnew,ALZOUBI2002459,Nanubolu2015}. These differences reflect the difficulty to control the experimental setup for a wide range of frequencies and to synthesize a pure sample especially in the case of those polymorphic crystals, which undergo phase transitions under specific thermodynamic conditions. In particular, as explained in Ref.~\cite{Agnew}, the crystallization of paracetamol in form II is often not perfect, as some traces of metacetamol may remain present, leading to partially mixed Raman spectra. 

\section{Conclusions \label{sec:conclusions}}
In this paper, we derived and implemented a real-space formulation of perturbation theory for homogeneous electric fields within an all-electron, numeric atom-centered orbitals DFT framework. We validated the approach by computing polarizabilities (and dielectric constants) of molecules and solids. In particular, we have
shown that these calculations can be systematically converged with respect to the numerical parameters used in the computation. Due to the slow convergence of polarizabilities with respect to the basis set size for isolated systems, we propose a simple solution based on the addition of so-called ``ghost" atoms (i.e. only basis functions) in parts of space that are not densely populated. Also, we show how to stabilize our implementation for situations where small differences between occupied and unoccupied eigenvalues are present, arriving at a formulation which proved always stable.
The scaling behavior of our implementation for the calculation of polarizabilities is between $O(N)$ and $O(N^{2})$ for both non-periodic~($O(N^{1.3})$) and periodic systems~($O(N^{1.4})$). In order to reduce the total time to $O(N)$, more advanced algorithms~\cite{Ochsenfeld-1997,Liang-2005} for the evaluation of the density-matrix response $\vec{P}^{(1)}$ could be pursued in the future.

We have tested our approach for the computation of dielectric constants by comparing theoretical and experimental literature data for a variety of semiconductors, obtaining very good agreement.
To highlight the power of our PT implementation, we applied it to the calculation of anharmonic Raman spectra of the isolated molecule of paracetamol, as well as two of its polymorphic crystal forms, which involved the computation of hundreds of thousands of polarizability tensors in order to build the time series. 
We obtained good agreement with experiment in all cases especially for the lineshapes, which highlights the power of \textit{ab initio} MD to capture anharmonic phonon frequencies and lifetimes, as well as the respective material properties~\cite{Carbogno:2017gc}. Regarding the calculated peak positions, we observe blue-shifts in the NH and CH stretching regions that stem from the lack of nuclear quantum effects in the MD simulations, as we explicitly show for the isolated molecule. We also found that these spectra are very sensitive to the 
xc-functional employed for the assessment of the potential-energy surface, but that they are rather insensitive to the xc-functional employed for the calculation of the polarizabilities. In fact, we obtain correct spectra in a computationally efficient manner by using LDA for the polarizability tensors, but the PBE functional with many-body van der Waals corrections for the PES. We have shown that having such an efficient implementation that gives access to anharmonic Raman signals
will be extremely useful for the analysis of experimental Raman spectra, which are often used to characterize new polymorphic forms of (molecular) crystals. 

The data presented in this work as well as the input and output files used to produce it are publicly available as a dataset~\cite{nomad-repository} on the NOMAD Repository.

\ack
H.S. acknowledges Vivekanand Gobre and Wanzhen Liang for inspiring discussions. M.R. and N.R. acknowledge computer time from the Swiss National Supercomputing Centre (CSCS), under Project No. s786. We further acknowledge Volker Blum for his continued support during this project, Raul Laasner for providing the first version of the Pulay mixer we use, and Florian Knoop for critical proof-reading. The project received funding from the Einstein foundation (project ETERNAL), the Deutsche Forschungsgemeinschaft (DFG) through SFB 951, and the European Union’s Horizon 2020 research and innovation program under grant agreement no. 676580 with The Novel Materials Discovery (NOMAD) Laboratory, a European Center of Excellence. P.R. acknowledges financial support from the Academy of Finland through its Centres of Excellence Program (Project No. 251748 and 284621).

\appendix

\section{Validation of the Polarisability Tensor for Molecules}
\label{sec:val}
To validate our implementation for isolated systems, we compared the PT polarizabilities of 16~dimers~(see Tab.~\ref{appendix_tab:dimers}), 5~trimers~(see Tab.~\ref{appendix_tab:trimers}), and 11~molecules~(see Tab.~\ref{appendix_tab:mol}) with those obtained by finite differences. In the latter case, the polarizability tensors were calculated using a finite, external electric field perurbation of~$\pm$~0.01~{V/\AA}. All calculations were performed for fully relaxed geometries~(remaining maximum force components smaller than $10^{-4}$~eV/$\mbox{\AA}$) at the LDA level of theory using \textit{tier 2} basis sets and ``really tight'' defaults for all other numerical parameters such as integration grids. 
In all cases, we find that the observed deviations between the polarizabilities obtained via PT and via finite differences are orders of magnitude smaller than the polarizabilities themselves, as also substantiated by the respective mean absolute errors~(MAE) and mean absolute percentage errors~(MAPE) given in Tabs.~\ref{appendix_tab:dimers}-\ref{appendix_tab:mol}. Please note that even the largest observed absolute error~(0.0018~$a_0^{3}$ for HCN) corresponds to a  very small relative error of only~$\sim 0.008$~\%.

\begin{table}
\begin{tabular}{lr|rrr} \hline \hline
     &      & PT~($a_0^3$)  &   AE$\cdot10^3$~($a_0^3$) &  APE$\cdot10^3$~(\%) \\\hline 
Cl$_2$   &   $\alpha_{xx}$  &   -24.123 &     0.07 &     0.29 \\ 
         &   $\alpha_{yy}$  &   -24.123 &     0.07 &     0.29 \\ 
         &   $\alpha_{zz}$  &   -41.309 &     0.10 &     0.24 \\ 
ClF      &   $\alpha_{xx}$  &   -15.927 &     0.05 &     0.31 \\ 
         &   $\alpha_{yy}$  &   -15.927 &     0.05 &     0.31 \\ 
         &   $\alpha_{zz}$  &   -22.292 &     0.06 &     0.27 \\ 
CO       &   $\alpha_{xx}$  &   -11.656 &     0.00 &     0.00 \\ 
         &   $\alpha_{yy}$  &   -11.656 &     0.00 &     0.00 \\ 
         &   $\alpha_{zz}$  &   -15.493 &     0.01 &     0.06 \\ 
CS       &   $\alpha_{xx}$  &   -22.239 &     0.03 &     0.13 \\ 
         &   $\alpha_{yy}$  &   -22.239 &     0.03 &     0.13 \\ 
         &   $\alpha_{zz}$  &   -37.652 &     0.07 &     0.19 \\ 
F$_2$    &   $\alpha_{xx}$  &    -6.170 &     0.11 &     1.78 \\ 
         &   $\alpha_{yy}$  &    -6.170 &     0.11 &     1.78 \\ 
         &   $\alpha_{zz}$  &   -11.684 &     0.16 &     1.37 \\ 
H$_2$    &   $\alpha_{xx}$  &    -3.902 &     0.01 &     0.26 \\ 
         &   $\alpha_{yy}$  &    -3.902 &     0.01 &     0.26 \\ 
         &   $\alpha_{zz}$  &    -7.532 &     0.03 &     0.40 \\ 
HCl      &   $\alpha_{xx}$  &   -16.815 &     0.00 &     0.00 \\ 
         &   $\alpha_{yy}$  &   -16.815 &     0.00 &     0.00 \\ 
         &   $\alpha_{zz}$  &   -18.868 &     0.05 &     0.26 \\ 
HF       &   $\alpha_{xx}$  &    -4.964 &     0.08 &     1.61 \\ 
         &   $\alpha_{yy}$  &    -4.964 &     0.08 &     1.61 \\ 
         &   $\alpha_{zz}$  &    -6.410 &     0.08 &     1.25 \\ 
Li$_2$   &   $\alpha_{xx}$  &  -120.631 &     1.16 &     0.96 \\ 
         &   $\alpha_{yy}$  &  -120.631 &     1.16 &     0.96 \\ 
         &   $\alpha_{zz}$  &  -231.987 &     3.83 &     1.65 \\ 
LiF      &   $\alpha_{xx}$  &   -11.162 &     0.19 &     1.70 \\ 
         &   $\alpha_{yy}$  &   -11.162 &     0.19 &     1.70 \\ 
         &   $\alpha_{zz}$  &   -11.064 &     0.06 &     0.54 \\ 
LiH      &   $\alpha_{xx}$  &   -29.868 &     0.43 &     1.44 \\ 
         &   $\alpha_{yy}$  &   -29.868 &     0.43 &     1.44 \\ 
         &   $\alpha_{zz}$  &   -30.634 &     0.97 &     3.17 \\\hline\hline
\end{tabular}
\caption{Polarizability tensor elements~$\alpha_{ii}$ for 16 dimers, as computed with the presented PT implementation at the LDA level of theory. Additionally, absolute errors~(AE) and absolute percentage erros~(APE) with respect to finite difference calculations are given. Please note that this errors are several orders of magnitude smaller than the relevant digits in~$\alpha_{ii}$.}
\label{appendix_tab:dimers}
\end{table}
 
\begin{table}
\begin{tabular}{lr|rrr} \hline \hline
     &      & PT~($a_0^3$)  &   AE$\cdot10^3$~($a_0^3$) &  APE$\cdot10^3$~(\%) \\\hline 
N$_2$    &   $\alpha_{xx}$  &    -9.923 &     0.04 &     0.40 \\ 
         &   $\alpha_{yy}$  &    -9.923 &     0.04 &     0.40 \\ 
         &   $\alpha_{zz}$  &   -15.033 &     0.14 &     0.93 \\ 
Na$_2$   &   $\alpha_{xx}$  &  -121.132 &     0.58 &     0.48 \\ 
         &   $\alpha_{yy}$  &  -121.132 &     0.58 &     0.48 \\ 
         &   $\alpha_{zz}$  &  -283.915 &     7.22 &     2.54 \\ 
NaCl     &   $\alpha_{xx}$  &   -28.156 &     0.22 &     0.78 \\ 
         &   $\alpha_{yy}$  &   -28.156 &     0.22 &     0.78 \\ 
         &   $\alpha_{zz}$  &   -40.558 &     0.00 &     0.00 \\ 
P$_2$    &   $\alpha_{xx}$  &   -34.724 &     0.02 &     0.06 \\ 
         &   $\alpha_{yy}$  &   -34.724 &     0.02 &     0.06 \\ 
         &   $\alpha_{zz}$  &   -67.280 &     0.09 &     0.13 \\ 
SiO      &   $\alpha_{xx}$  &   -24.570 &     0.40 &     1.63 \\ 
         &   $\alpha_{yy}$  &   -24.570 &     0.40 &     1.63 \\
         &   $\alpha_{zz}$  &   -34.021 &     0.13 &     0.38 \\\hline
Mean     &                  &           &     0.41 &     0.77 \\\hline \hline
\end{tabular}
\caption{(continued) Polarizability tensor elements~$\alpha_{ii}$ for 16 dimers, as computed with the presented PT implementation at the LDA level of theory. Additionally, absolute errors~(AE) and absolute percentage erros~(APE) with respect to finite difference calculations are given. Please note that this errors are several orders of magnitude smaller than the relevant digits in~$\alpha_{ii}$.}
\label{appendix_tab2:dimers}
\end{table}

\begin{table}
\begin{tabular}{lr|rrr} \hline \hline
     &      & PT~($a_0^3$)  &   AE$\cdot10^3$~($a_0^3$) &  APE$\cdot10^3$~(\%) \\\hline 
CO$_2$   &   $\alpha_{xx}$  &   -12.041 &     0.14 &     1.16 \\ 
         &   $\alpha_{yy}$  &   -12.041 &     0.14 &     1.16 \\ 
         &   $\alpha_{zz}$  &   -26.559 &     0.23 &     0.87 \\ 
H$_2$O   &   $\alpha_{xx}$  &    -8.576 &     0.06 &     0.70 \\ 
         &   $\alpha_{yy}$  &    -9.795 &     0.11 &     1.12 \\ 
         &   $\alpha_{zz}$  &    -9.191 &     0.03 &     0.33 \\ 
HCN      &   $\alpha_{xx}$  &   -13.101 &     0.08 &     0.61 \\ 
         &   $\alpha_{yy}$  &   -13.101 &     0.08 &     0.61 \\ 
         &   $\alpha_{zz}$  &   -23.102 &     1.88 &     8.14 \\ 
SH$_2$   &   $\alpha_{xx}$  &   -23.169 &     0.03 &     0.13 \\ 
         &   $\alpha_{yy}$  &   -24.109 &     0.24 &     1.00 \\ 
         &   $\alpha_{zz}$  &   -24.052 &     0.04 &     0.17 \\ 
SO$_2$   &   $\alpha_{xx}$  &   -18.869 &     0.03 &     0.16 \\ 
         &   $\alpha_{yy}$  &   -33.634 &     0.05 &     0.15 \\ 
         &   $\alpha_{zz}$  &   -22.710 &     0.01 &     0.04 \\ \hline
Mean     &                  &           &     0.21 &     1.09  \% \\
\hline \hline
\end{tabular}
\caption{Polarizability tensor elements~$\alpha_{ii}$ for five trimers, as computed with the presented PT implementation at the LDA level of theory. Additionally, absolute errors~(AE) and absolute percentage erros~(APE) with respect to finite difference calculations are given. Please note that this errors are several orders of magnitude smaller than the relevant digits in~$\alpha_{ii}$.}
\label{appendix_tab:trimers}
\end{table}

\begin{table}
\begin{tabular}{lr|rrr} \hline \hline
     &      & PT~($a_0^3$)  &   AE$\cdot10^3$~($a_0^3$) &  APE$\cdot10^3$~(\%) \\\hline 
C$_2$H$_2$    &   $\alpha_{xx}$  &   -16.323 &     0.09 &     0.55 \\ 
              &   $\alpha_{yy}$  &   -16.323 &     0.09 &     0.55 \\ 
              &   $\alpha_{zz}$  &   -31.802 &     0.22 &     0.69 \\ 
C$_2$H$_4$    &   $\alpha_{xx}$  &   -20.208 &     0.10 &     0.49 \\ 
              &   $\alpha_{yy}$  &   -24.666 &     0.34 &     1.38 \\ 
              &   $\alpha_{zz}$  &   -35.705 &     0.11 &     0.31 \\ 
CH$_3$Cl      &   $\alpha_{xx}$  &   -26.327 &     0.02 &     0.08 \\ 
              &   $\alpha_{yy}$  &   -26.327 &     0.03 &     0.11 \\ 
              &   $\alpha_{zz}$  &   -35.999 &     0.10 &     0.28 \\ 
CH$_4$        &   $\alpha_{xx}$  &   -16.974 &     0.62 &     3.65 \\ 
              &   $\alpha_{yy}$  &   -16.974 &     0.62 &     3.65 \\ 
              &   $\alpha_{zz}$  &   -16.974 &     0.62 &     3.65 \\ 
H$_2$CO       &   $\alpha_{xx}$  &   -11.993 &     0.04 &     0.33 \\ 
              &   $\alpha_{yy}$  &   -18.333 &     0.09 &     0.49 \\ 
              &   $\alpha_{zz}$  &   -23.032 &     0.11 &     0.48 \\ 
H$_2$O$_2$    &   $\alpha_{xx}$  &   -13.596 &     0.12 &     0.88 \\ 
              &   $\alpha_{yy}$  &   -17.595 &     0.17 &     0.97 \\ 
              &   $\alpha_{zz}$  &   -12.362 &     0.13 &     1.05 \\ 
N$_2$H$_4$    &   $\alpha_{xx}$  &   -20.997 &     0.08 &     0.38 \\ 
              &   $\alpha_{yy}$  &   -25.882 &     0.07 &     0.27 \\ 
              &   $\alpha_{zz}$  &   -21.209 &     0.07 &     0.33 \\ 
NH$_3$        &   $\alpha_{xx}$  &   -13.339 &     0.02 &     0.15 \\ 
              &   $\alpha_{yy}$  &   -13.339 &     0.00 &     0.00 \\ 
              &   $\alpha_{zz}$  &   -14.608 &     0.07 &     0.48 \\ 
PH$_3$        &   $\alpha_{xx}$  &   -30.003 &     0.70 &     2.33 \\ 
              &   $\alpha_{yy}$  &   -30.003 &     0.70 &     2.33 \\ 
              &   $\alpha_{zz}$  &   -31.116 &     0.01 &     0.03 \\ 
Si$_2$H$_6$   &   $\alpha_{xx}$  &   -57.444 &     0.20 &     0.35 \\ 
              &   $\alpha_{yy}$  &   -57.444 &     0.23 &     0.40 \\ 
              &   $\alpha_{zz}$  &   -77.035 &     0.36 &     0.47 \\ 
SiH$_4$       &   $\alpha_{xx}$  &   -31.967 &     0.14 &     0.44 \\ 
              &   $\alpha_{yy}$  &   -31.967 &     0.14 &     0.44 \\ 
              &   $\alpha_{zz}$  &   -31.967 &     0.14 &     0.44 \\\hline
Mean          &                  &           &     0.20 &     0.86 \\\hline \hline
\end{tabular}
\caption{Polarizability tensor elements~$\alpha_{ii}$ for eleven molecules, as computed with the presented PT implementation at the LDA level of theory. Additionally, absolute errors~(AE) and absolute percentage erros~(APE) with respect to finite difference calculations are given. Please note that this errors are several orders of magnitude smaller than the relevant digits in~$\alpha_{ii}$.}
\label{appendix_tab:mol}
\end{table}


\begin{thebibliography}{10}
\expandafter\ifx\csname url\endcsname\relax
  \def\url#1{{\tt #1}}\fi
\expandafter\ifx\csname urlprefix\endcsname\relax\def\urlprefix{URL }\fi
\providecommand{\eprint}[2][]{\url{#2}}

\bibitem{DalCorso1996}
Corso A~D, Mauri F and Rubio A 1996 {\em Phys. Rev. B\/} {\bf 53} 15638--15642
  \urlprefix\url{http://link.aps.org/doi/10.1103/PhysRevB.53.15638}

\bibitem{Andrade:2007kg}
Andrade X, Botti S, Marques M~A~L and Rubio A 2007 {\em J. Chem. Phys.\/} {\bf 126} 184106

\bibitem{Gonze1997-1}
Gonze X 1997 {\em Phys. Rev. B\/} {\bf 55} 10337--10354 
  \urlprefix\url{http://link.aps.org/doi/10.1103/PhysRevB.55.10337}

\bibitem{Gonze1997-2}
Gonze X and Lee C 1997 {\em Phys. Rev. B\/} {\bf 55} 10355--10368
  \urlprefix\url{http://link.aps.org/doi/10.1103/PhysRevB.55.10355}

\bibitem{PutrinoParrinelloDFPT2000}
Putrino A, Sebastiani D and Parrinello M 2000 {\em J. Chem. Phys.\/} {\bf 113}
  7102--7109 \urlprefix\url{http://dx.doi.org/10.1063/1.1312830}

\bibitem{Baroni-2001}
Baroni S, de~Gironcoli S, {Dal Corso} A and Giannozzi P 2001 {\em Rev. Mod.
  Phys.\/} {\bf 73} 515--562 
  \urlprefix\url{http://link.aps.org/doi/10.1103/RevModPhys.73.515}

\bibitem{Gerratt-1967}
Gerratt J and Mills I~M 1968 {\em J. Chem. Phys.\/} {\bf 49} 1719--1729  \urlprefix\url{http://link.aip.org/link/?JCP/49/1719/1
  http://aip.scitation.org/doi/10.1063/1.1670299}

\bibitem{Pople-1979}
Pople J~A, Krishnan R, Schlegel H~B and Binkley J~S 1979 {\em Int. J. Quantum
  Chem.\/} {\bf 16} 225--241 
  \urlprefix\url{http://dx.doi.org/10.1002/qua.560160825}

\bibitem{Dykstra-1984}
Dykstra C~E and Jasien P~G 1984 {\em Chem. Phys. Lett.\/} {\bf 109} 388--393
  
  \urlprefix\url{http://www.sciencedirect.com/science/article/pii/0009261484856079}

\bibitem{Frisch-1990}
Frisch M, Head-Gordon M and Pople J 1990 {\em Chem. Phys.\/} {\bf 141} 189--196
  
  \urlprefix\url{http://www.sciencedirect.com/science/article/pii/030101049087055G}

\bibitem{Ochsenfeld-1997}
Ochsenfeld C and Head-Gordon M 1997 {\em Chem. Phys. Lett.\/} {\bf 270}
  399--405 
  \urlprefix\url{http://www.sciencedirect.com/science/article/pii/S0009261497004028}

\bibitem{Liang-2005}
Liang W, Zhao Y and Head-Gordon M 2005 {\em J. Chem. Phys.\/} {\bf 123} 194106
  \urlprefix\url{http://link.aip.org/link/?JCP/123/194106/1
  http://scitation.aip.org/content/aip/journal/jcp/123/19/10.1063/1.2114847}

\bibitem{Blum2009}
Blum V, Gehrke R, Hanke F, Havu P, Havu V, Ren X, Reuter K and Scheffler M 2009
  {\em Comput. Phys. Commun.\/} {\bf 180} 2175--2196  
  \urlprefix\url{http://linkinghub.elsevier.com/retrieve/pii/S0010465509002033}

\bibitem{Havu/etal:2009}
Havu V, Blum V, Havu P and Scheffler M 2009 {\em J. Comput. Phys.\/} {\bf 228}
  8367--8379 
  \urlprefix\url{http://linkinghub.elsevier.com/retrieve/pii/S0021999109004458}

\bibitem{Ren/etal:2012}
Ren X, Rinke P, Blum V, Wieferink J, Tkatchenko A, Sanfilippo A, Reuter K and
  Scheffler M 2012 {\em New J. Phys.\/} {\bf 14} 53020 
  \urlprefix\url{http://stacks.iop.org/1367-2630/14/i=5/a=053020?key=crossref.351b343783c2c1df1596219a941a74eb}

\bibitem{Shang:2017hn}
Shang H, Carbogno C, Rinke P and Scheffler M 2017 {\em Comput. Phys. Commun.\/}
  {\bf 215} 26--46

\bibitem{Laasner:fx}
Laasner R, Huhn W, Colell J, Theis T, Yu V~W, Warren W and Blum V 2018  {\bf
  arXiv:1805.12225}

\bibitem{Putrino_PRL_2002}
Putrino A and Parrinello M 2002 {\em Phys. Rev. Lett.\/} {\bf 88} 176401
  \urlprefix\url{https://link.aps.org/doi/10.1103/PhysRevLett.88.176401}

\bibitem{Pagliai_JCP2008}
Pagliai M, Cavazzoni C, Cardini G, Erbacci G, Parrinello M and Schettino V 2008
  {\em J. Chem. Phys.\/} {\bf 128} 224514
  \urlprefix\url{http://dx.doi.org/10.1063/1.2936988}

\bibitem{Gonze-1989}
Gonze X and Vigneron J~P 1989 {\em Phys. Rev. B\/} {\bf 39} 13120--13128
  \urlprefix\url{http://link.aps.org/doi/10.1103/PhysRevB.39.13120}

\bibitem{Mosley1993}
Mosley C~D~H and Fripiat J~G 1993 {\em Int. J. Quant. Chem.\/} {\bf 46}

\bibitem{Giannozzi1991}
Giannozzi P, de~Gironcoli S, Pavone P and Baroni S 1991 {\em Phys. Rev. B\/}
  {\bf 43} 7231--7242
  \urlprefix\url{http://link.aps.org/doi/10.1103/PhysRevB.43.7231}

\bibitem{Ferrero2008}
Ferrero M, R{\'{e}}rat M, Orlando R and Dovesi R 2008 {\em J. Comput. Chem.\/}
  {\bf 29} 1450--1459 \urlprefix\url{http://dx.doi.org/10.1002/jcc.20905}

\bibitem{Knuth:2015kc}
Knuth F, Carbogno C, Atalla V, Blum V and Scheffler M 2015 {\em Comput. Phys.
  Commun.\/} {\bf 190} 33--50

\bibitem{Griffiths:1999ud}
Griffiths D~J 1999 {\em {Introduction To Electrodynamics}\/} 3rd ed (Prentice
  Hall)

\bibitem{ashcroft}
Ashcroft N~W and Mermin D~N 1976 {\em {Solid State Physics}\/} 1st ed (Toronto:
  Thomson Learning) ISBN 0030839939
  \urlprefix\url{http://www.amazon.com/exec/obidos/redirect?tag=citeulike07-20{\&}path=ASIN/0030839939}

\bibitem{Vanderbilt1993}
King-Smith R~D and Vanderbilt D 1993 {\em Phys. Rev. B\/} {\bf 47} 1651--1654
  \urlprefix\url{http://link.aps.org/doi/10.1103/PhysRevB.47.1651}

\bibitem{Resta1994}
Resta R 1994 {\em Rev. Mod. Phys.\/} {\bf 66} 899--915
  \urlprefix\url{http://link.aps.org/doi/10.1103/RevModPhys.66.899}

\bibitem{Spaldin:2012ez}
Spaldin N~A 2012 {\em J. Solid State Chem.\/} {\bf 195} 2--10

\bibitem{Pulay1980}
Pulay P 1980 {\em Chem. Phys. Lett.\/} {\bf 73} 393--398 
  \urlprefix\url{http://www.sciencedirect.com/science/article/pii/0009261480803964}

\bibitem{Delley1990}
Delley B 1990 {\em J. Chem. Phys.\/} {\bf 92} 508 
  \urlprefix\url{http://scitation.aip.org/content/aip/journal/jcp/92/1/10.1063/1.458452}

\bibitem{Perdew/Zunger:1981}
Perdew J and Zunger A 1981 {\em Phys.\ Rev.\ B\/} {\bf 23} 5048

\bibitem{Ceperley/Alder:1980}
Ceperley D~M and Alder B~J 1980 {\em Phys. Rev. Lett.\/} {\bf 45} 566

\bibitem{Perdew-1}
Perdew J~P, Burke K and Ernzerhof M 1996 {\em Phys. Rev. Lett.\/} {\bf 77}
  3865--3868
  \urlprefix\url{https://link.aps.org/doi/10.1103/PhysRevLett.77.3865}

\bibitem{Perdew-2}
Perdew J~P, Burke K and Ernzerhof M 1997 {\em Phys. Rev. Lett.\/} {\bf 78} 1396
  \urlprefix\url{https://link.aps.org/doi/10.1103/PhysRevLett.78.1396}

\bibitem{Marques:2012bu}
Marques M~A~L, Oliveira M~J~T and Burnus T 2012 {\em Comp. Phys. Comm.\/} {\bf
  183} 2272--2281

\bibitem{Levchenko2015}
Levchenko S~V, Ren X, Wieferink J, Johanni R, Rinke P, Blum V and Scheffler M
  2015 {\em Comput. Phys. Commun.\/} {\bf 192} 60--69 
  \urlprefix\url{http://linkinghub.elsevier.com/retrieve/pii/S001046551500079X}

\bibitem{Johnson1994}
Johnson B~G and Fisch M~J 1994 {\em J. Chem. Phys.\/} {\bf 100} 7429 
  \urlprefix\url{http://scitation.aip.org/content/aip/journal/jcp/100/10/10.1063/1.466887}

\bibitem{Gironcoli_PRB_1995}
de~Gironcoli S 1995 {\em Phys. Rev. B\/} {\bf 51} 6773--6776
  \urlprefix\url{https://link.aps.org/doi/10.1103/PhysRevB.51.6773}

\bibitem{Rappoport_Furche_JCP2010}
Rappoport D and Furche F 2010 {\em J. Chem. Phys.\/} {\bf 133} 134105
  \urlprefix\url{http://dx.doi.org/10.1063/1.3484283}

\bibitem{hyperpolarizability_basis_convergence}
Vila F~D, Strubbe D~A, Takimoto Y, Andrade X, Rubio A, Louie S~G and Rehr J~J
  2010 {\em J. Chem. Phys.\/} {\bf 133} 34111
  \urlprefix\url{http://dx.doi.org/10.1063/1.3457362}

\bibitem{perdew-1992}
Perdew J~P and Wang Y 1992 {\em Phys. Rev. B\/} {\bf 45} 13244--13249

\bibitem{Gajdo2006}
Gajdo\ifmmode~\check{s}\else \v{s}\fi{} M, Hummer K, Kresse G, Furthm\"uller J
  and Bechstedt F 2006 {\em Phys. Rev. B\/} {\bf 73}(4) 045112
  \urlprefix\url{https://link.aps.org/doi/10.1103/PhysRevB.73.045112}

\bibitem{Petousis2016}
Petousis I, Chen W, Hautier G, Graf T, Schladt T~D, Persson K~A and Prinz F~B
  2016 {\em Phys. Rev. B\/} {\bf 93} 115151
  \urlprefix\url{https://link.aps.org/doi/10.1103/PhysRevB.93.115151}

\bibitem{Gironcoli1989}
de~Gironcoli S, Baroni S and Resta R 1989 {\em Phys. Rev. Lett.\/} {\bf 62}
  2853--2856
  \urlprefix\url{https://link.aps.org/doi/10.1103/PhysRevLett.62.2853}

\bibitem{Louie1982}
Louie S~G, Froyen S and Cohen M~L 1982 {\em Phys. Rev. B\/} {\bf 26} 1738--1742
  \urlprefix\url{https://link.aps.org/doi/10.1103/PhysRevB.26.1738}

\bibitem{DalCorso1993}
{Dal Corso} A, Baroni S, Resta R and de~Gironcoli S 1993 {\em Phys. Rev. B\/}
  {\bf 47} 3588--3592
  \urlprefix\url{https://link.aps.org/doi/10.1103/PhysRevB.47.3588}

\bibitem{Philipp1963-Si-GaP}
Philipp H~R and Ehrenreich H 1963 {\em Phys. Rev.\/} {\bf 129}(4) 1550--1560
  \urlprefix\url{https://link.aps.org/doi/10.1103/PhysRev.129.1550}

\bibitem{Beer1968-AlP-1}
Beer S~Z, Jackovitz J~F, Feldman D~W and Parker J~H 1968 {\em Phys. Lett. A\/}
  {\bf 26} 331--332 
  \urlprefix\url{http://www.sciencedirect.com/science/article/pii/0375960168906804}

\bibitem{Monemar1970-AlP-2}
Monemar B 1970 {\em Solid State Commun.\/} {\bf 8} 1295--1298 
  \urlprefix\url{http://www.sciencedirect.com/science/article/pii/003810987090623X}

\bibitem{Fern1971-AlAs}
Fern R~E and Onton A 1971 {\em J. Appl. Phys.\/} {\bf 42} 3499--3500
  \urlprefix\url{https://doi.org/10.1063/1.1660760}

\bibitem{Hass1962-AlSb-GaSb}
Hass M and Henvis B~W 1962 {\em J. Phys. Chem. Solids\/} {\bf 23} 1099--1104
  \urlprefix\url{http://www.sciencedirect.com/science/article/pii/0022369762901270}

\bibitem{Mariana_PRL_2016}
Rossi M, Gasparotto P and Ceriotti M 2016 {\em Phys. Rev. Lett.\/} {\bf 117}
  115702
  \urlprefix\url{https://link.aps.org/doi/10.1103/PhysRevLett.117.115702}

\bibitem{Warshel1971}
Warshel A 1971 {\em J. Chem. Phys.\/} {\bf 54} 5324--5330

\bibitem{Neugebauer_JCC2002}
Neugebauer J, Reiher M, Kind C and Hess B~A 2002 {\em J. Comput. Chem.\/} {\bf
  23} 895--910 
  \urlprefix\url{http://dx.doi.org/10.1002/jcc.10089}

\bibitem{Veithen:2005gf}
Veithen M, Gonze X and Ghosez P 2005 {\em Phys. Rev. B\/} {\bf 71} 125107

\bibitem{Berne_Pecora}
Ross-Murphy S~B 1977 {\em Br. Polym. J.\/} {\bf 9} 177  
  \urlprefix\url{http://dx.doi.org/10.1002/pi.4980090216}

\bibitem{Ramirez_2004}
Ram\'irez R, L{\'{o}}pez-Ciudad T, P P~K and Marx D 2004 {\em J. Chem.
  Phys.\/} {\bf 121} 3973--3983
  \urlprefix\url{https://doi.org/10.1063/1.1774986}

\bibitem{McCreery_2005}
McCreery R~L 2005 {\em {Magnitude of Raman Scattering}\/} (John Wiley {\&}
  Sons.) pp 15--33 ISBN 9780471721642
  \urlprefix\url{http://dx.doi.org/10.1002/0471721646.ch2}

\bibitem{Keresztury_1993}
Keresztury G, Holly S, Besenyei G, Varga J, Wang A and Durig J~R 1993 {\em
  Spectrochim. Acta Part A Mol. Spectrosc.\/} {\bf 49} 2007--2026 
  \urlprefix\url{http://www.sciencedirect.com/science/article/pii/S0584853909910121}

\bibitem{Chalmers-Griffiths_2001}
Chalmers J~M and Griffiths P~R (eds) 2001 {\em {Handbook of Vibrational
  Spectroscopy}\/} (John Wiley {\&} Sons, Ltd)
  \urlprefix\url{https://doi.org/10.1002/0470027320}

\bibitem{Distasio_Tkatchenko_2014}
Distasio R~A, Gobre V~V and Tkatchenko A 2014 {\em J. Phys. Condens. Matter\/} {\bf
  26} 213202 \urlprefix\url{http://stacks.iop.org/0953-8984/26/i=21/a=213202}

\bibitem{ReillyTkatchen2014}
Reilly A~M and Tkatchenko A 2014 {\em Phys. Rev. Lett.\/} {\bf 113} 55701
  \urlprefix\url{https://link.aps.org/doi/10.1103/PhysRevLett.113.055701}

\bibitem{HojaTkatchen2017}
Hoja J, Reilly A~M and Tkatchenko A 2017 {\em Wiley Interdiscip. Rev. Comput.
  Mol. Sci.\/} {\bf 7} e1294----n/a 
  \urlprefix\url{http://dx.doi.org/10.1002/wcms.1294}

\bibitem{RossiBlum2014}
Rossi M, Chutia S, Scheffler M and Blum V 2014 {\em J. Phys. Chem. A\/} {\bf
  118} 7349--7359 \urlprefix\url{http://dx.doi.org/10.1021/jp412055r}

\bibitem{BaldaufRossi2015}
Baldauf C and Rossi M 2015 {\em J. Phys. Condens. Matter\/} {\bf 27} 493002
  \urlprefix\url{http://stacks.iop.org/0953-8984/27/i=49/a=493002}

\bibitem{RossiMano2014}
Rossi M, Ceriotti M and Manolopoulos D~E 2014 {\em J. Chem. Phys.\/} {\bf 140}
  234116 \urlprefix\url{http://dx.doi.org/10.1063/1.4883861}

\bibitem{RossiCeriotti2018}
Rossi M, Kapil V and Ceriotti M 2017 {\em J. Chem. Phys.\/}
  {\bf 148} 102301

\bibitem{RossiMichaelides2015}
Rossi M, Fang W and Michaelides A 2015 {\em J. Phys. Chem. Lett.\/} {\bf 6}
  4233--4238 \urlprefix\url{http://dx.doi.org/10.1021/acs.jpclett.5b01899}

\bibitem{RossiCeriottiJCPcomm}
Rossi M, Liu H, Paesani F, Bowman J and Ceriotti M 2014 {\em J. Chem. Phys.\/} {\bf 141} 181101

\bibitem{Nanubolu}
Nanubolu J~B and Burley J~C 2012 {\em Mol. Pharm.\/} {\bf 9} 1544--1558
  \urlprefix\url{http://dx.doi.org/10.1021/mp300035g}

\bibitem{Agnew}
Agnew L~R, McGlone T, Wheatcroft H~P, Robertson A, Parsons A~R and Wilson C~C
  2017 {\em Cryst. Growth Des.\/} {\bf 17} 2418--2427
  \urlprefix\url{http://dx.doi.org/10.1021/acs.cgd.6b01831}

\bibitem{ALZOUBI2002459}
Al-Zoubi N, Koundourellis J~E and Malamataris S 2002 {\em J. Pharm. Biomed.
  Anal.\/} {\bf 29} 459--467 
  \urlprefix\url{http://www.sciencedirect.com/science/article/pii/S0731708502000985}

\bibitem{Nanubolu2015}
Nanubolu J~B and Burley J~C 2015 {\em CrystEngComm\/} {\bf 17} 5280--5287
  \urlprefix\url{http://dx.doi.org/10.1039/C5CE00008D}

\bibitem{Carbogno:2017gc}
Carbogono C, Ramprasad R, and Scheffler M 2017 {\em Phys. Rev. Lett.\/} {\bf 118} 175901
  \urlprefix\url{http://dx.doi.org/10.1103/PhysRevLett.118.175901}

\bibitem{nomad-repository}
Shang H, Raimbault N, Rinke P, Scheffler M, Rossi M, and Carbogno C, {\bf  DOI:} 10.17172/NOMAD/2018.02.16-1
  \urlprefix\url{http://dx.doi.org/10.17172/NOMAD/2018.02.16-1}

\end{thebibliography}

\providecommand{\newblock}{}

\end{document}